\documentclass[a4paper,fleqn,usenatbib]{mnras}

\usepackage[T1]{fontenc}
\usepackage{ae,aecompl}

\usepackage[utf8]{inputenc} 
\usepackage{graphicx}	
\usepackage{amsmath}	
\usepackage{amssymb}	
\usepackage{subcaption} 
\usepackage{xcolor}

\usepackage{newtxtext,newtxmath}

\newcommand\sualpha{\scriptsize\unslant\alpha\kern-0.075em}
\newcommand{\dif}{\mathrm{d}}

\title[AGN ionized outflows with Athena]{Spectral-timing of AGN ionized outflows with Athena} 

\author[A. Jur\'{a}\v{n}ov\'{a} et al.]{A. Jur\'{a}\v{n}ov\'{a}$^{1,2}$\thanks{E-mail: a.juranova@sron.nl},
	E. Costantini$^{1,2}$,
	P. Uttley$^{2}$
	\\
	$^{1}$SRON Netherlands Institute for Space Research, Niels Bohrweg 4, NL-2333 CA Leiden, the Netherlands\\
	$^{2}$Anton Pannekoek Institute, University of Amsterdam, Postbus 94249, NL-1090 GE Amsterdam, the Netherlands
}
\date{Accepted 2021 December 20. Received 2021 December 17; in original form 2021 October 10}

\pubyear{2022}

\begin{document}
	\label{firstpage}
	\pagerange{\pageref{firstpage}--\pageref{lastpage}}
	\maketitle
	
	\begin{abstract}
		Spectral-timing techniques have proven valuable in studying the interplay between the X-ray corona and the accretion disc in variable active galactic nuclei (AGN). Under certain conditions, photo-ionized outflows emerging from central AGN regions also play a role in the observable spectral-timing properties of the nuclear components. The variable ionizing flux causes the intervening gas to ionize or recombine, resulting in a time-dependent absorption spectrum. Understanding the spectral-timing properties of these outflows is critical for the determination of their role in the AGN environment, but also the correct interpretation of timing signatures of other AGN components. In this paper, we test the capabilities of the \textit{Athena} X-IFU instrument in studying the spectral and spectral-timing properties of a black hole system displaying a variable outflow. We take the narrow-line Seyfert~1 IRAS~13224-3809 as a test case. Our findings show that while the non-linear response of the absorbing medium can result in complex behaviour of time lags, the resulting decrease in the coherence can be used to constrain gas density and distance to the central source. Ultimately, modelling the coherence spectra of AGN outflows may constitute a valuable tool in studying the physical properties of the outflowing gas.
	\end{abstract}
	
	
	
	
	\section{Introduction}\label{sec:intro}
	
	The variability of AGN X-ray emission has been extensively used to study the properties of the innermost regions surrounding the central supermassive black hole. The time-scales of these flux variations range from years down to hundreds of seconds, with an amplitude reaching in some cases an order of magnitude \citep[e.g.][]{Ponti2012}.
	
	The short variability time-scale limits the number of X-ray photons received for time-resolved spectroscopy. Fourier spectral-timing techniques \citep[see][for a review]{Uttley2014} have proven invaluable for overcoming this limitation, as they rely on statistical properties of the analysed light curves, rather than a high signal-to-noise ratio in a given time and energy bin. Furthermore, they enable isolation and examination of processes occurring on different time-scales, provided they can be distinguished by the energy at which they are manifested. Notably, the use of Fourier techniques allows measurement of delays associated with light-travel time between individual sources of X-rays within the central region, facilitating studies of the geometry and dynamics of the innermost accretion flow \citep{Fabian2009, Zoghbi2012, DeMarco2013, Kara2019, Alston2020}.
	
	In the current understanding of the central environment of AGN, the source of the primary variable X-ray flux is a compact corona \citep{Galeev1979}, lying close above the accretion disc. There, the thermal photons from the accretion disc are scattered by electrons to higher energies through the inverse Compton process, forming a power-law spectrum \citep{Haardt1993}. Part of this emission then irradiates the accretion disk, giving rise to a complex reflection spectrum, consisting of emission lines as well as a continuum component \citep{Ross1993}. 
	The reflection process leads to time lags between the primary and secondary radiation due to the additional light travel time, an effect known as reverberation \citep{Stella1990, Reynolds1999, Uttley2014}.
	
	If gas outflows are present in such an environment, additional delays can be produced, either by reverberation on the outflowing material out of the line of sight \citep[e.g.][]{Miller2010, Mangham2017, Mizumoto2019}, or due to interaction of the photons with gas observed along the line of sight \citep{Silva2016}. The latter can occur when the ionization of the outflowing gas evolves with the incoming variable X-ray radiation \citep{Nicastro1999a}. 
	
	Indeed, absorption features associated with outflowing material are present in many AGN \citep[see][for a review]{Laha2021}. These outflows have been reported over a wide range of ionization, as well as column density and outflow velocity. Outflows of lower ionization (with ionization parameter $\log \xi \sim 0-2$, where $\xi$ is in units of $\rm erg~s^{-1}~cm $) are often referred to as warm absorbers and have column densities typically between $N_{\rm H} \sim 10^{20}-10^{22}~\rm cm^{-2}$ and outflow velocities of $100-1000~\rm km~s^{-1}$ \citep[e.g.][]{Reynolds1997,George1998,Blustin2005}. Absorbers of higher ionisation ($\log \xi \sim 3-6$) show a generally larger column density $N_{\rm H} \sim 10^{22}-10^{24}~\rm cm^{-2}$ and a higher velocity of $10^4-10^5~\rm km~s^{-1}$, for which they are also known as ultra-fast outflows \citep[e.g.][]{Tombesi2010, Cappi2013}.
	
	The X-ray spectral-timing effects of absorbing outflows depend on the gas density, and consequently the distance from the ionizing source, as demonstrated by \citet{Silva2016} on a multi-component warm absorber. Their results imply the possibility of determination of these properties from the observed time lags.
	
	Unfortunately, current instrumentation provides only limited possibilities for inferring gas properties from spectral-timing analysis of AGN outflows. Consequently, these methods can only be applied in the case of low-ionization gas with a relatively high opacity. However, the large effective area and high spectral resolution in the energy range $0.2-12~\rm keV$ offered by the X-ray Integral Field Unit \citep[X-IFU;][]{Barret2018} onboard the \textit{Athena} X-ray observatory \citep{Nandra2013} will enable detailed spectral timing studies of AGN structure, including highly ionized outflows. 
	
	For the timing analysis of AGN ionized outflows, narrow-line Seyfert 1 galaxies (NLS1), in which outflows are commonly found \citep{Leighly1997}, are particularly suitable sources to target. The central flux shows strong variability, which occurs on time-scales from days to hours and the amplitude change may exceed an order of magnitude in the X-ray flux \citep{Boller1996, Leighly1999, Komossa2000}.
	
	In this study, we aim to demonstrate the capabilities of Fourier spectral-timing analysis of AGN outflows, applied to future \textit{Athena} X-IFU observations. In particular, we simulate observations of ionized outflows in a highly variable NLS1, taking IRAS 13224-3809 as a model example, and discuss how the characteristic response of the absorption spectral features reflects in the time lags and the coherence extracted from the observed light curves. We demonstrate that the coherence can be used to constrain the density and distance of the absorbing medium. This, in turn, is essential information to understand the formation and the impact of these outflows in the context of AGN feedback.
	
	This paper is structured as follows. Section \ref{sec:simulations} is dedicated to a description of the simulations, and the timing analysis of the resulting light curves is presented in Section \ref{sec:timing}. There, in Section \ref{sec:3xi}, the behaviour of outflows in an AGN, described by a simple baseline SED, is presented. Our findings are then compared to a more complex situation, where the gas responds to an SED with a lagging component (Section \ref{sec:varSED}). We explore a new method for gas distance determination as well as its limitations and future prospects in Section \ref{sec:discussion} and summarize our conclusions in Section \ref{sec:conclusions}.
	
	\section{Simulations}\label{sec:simulations}
	
	In order to simulate the behaviour of the gas response, we generated a realistic, yet simply-parametrized spectral energy distribution, which was used to determine the ionization balance, and a light curve representative of a highly-variable NLS1. Using this modelled source behaviour, we simulated the evolution of ionization properties of outflows under different conditions to encompass a wide range of the gas properties. A detailed description of this process, including the final simulation of \textit{Athena} X-IFU observations, is the main subject of this section.

	\subsection{SED parameters}\label{sec:params}

	To construct a realistic source of the ionizing radiation, we take the well-observed IRAS~13224-3809 as a model example. This AGN is a typical NLS1 located at $ z = 0.0658 $, with high X-ray variability \citep{Gallo2004} and a presence of variable blueshifted absorption lines ascribed to a highly ionized gas, possibly outflowing at $\sim 0.2~c$ \citep{Parker2017, Chartas2018, Jiang2018}.
	We base our simulations on the data obtained during the \textit{XMM-Newton} observing campaigns conducted in 2011 and 2016 (PI: A. C. Fabian), totalling over 1.5 Ms of data. 
	The data were reduced with the standard procedures of the \textit{XMM--Newton} Science Analysis System version 18.0.0. We accounted for events received by EPIC-pn during readout (out-of-time events) and filtered out the time periods affected by the so-called soft-proton flares by excluding the data where the count rate deviated from the mean by more than 3~$\sigma$ (3-$\sigma$ clipping). Short exposure losses (up to 1 ks) resulting from this procedure were filled in the light curve by linear interpolation and applying appropriate Poisson noise. 
	
	For the purpose of this study, only EPIC-pn light curves were used to obtain an average power spectrum, constructed from 100-ks segments with a $500~\rm s$ time step. Consequently, the range of frequencies covered is $ 10^{-5} - 10^{-3}$ Hz. 
	
\begin{figure}
		\centering
		\includegraphics[width=\columnwidth]{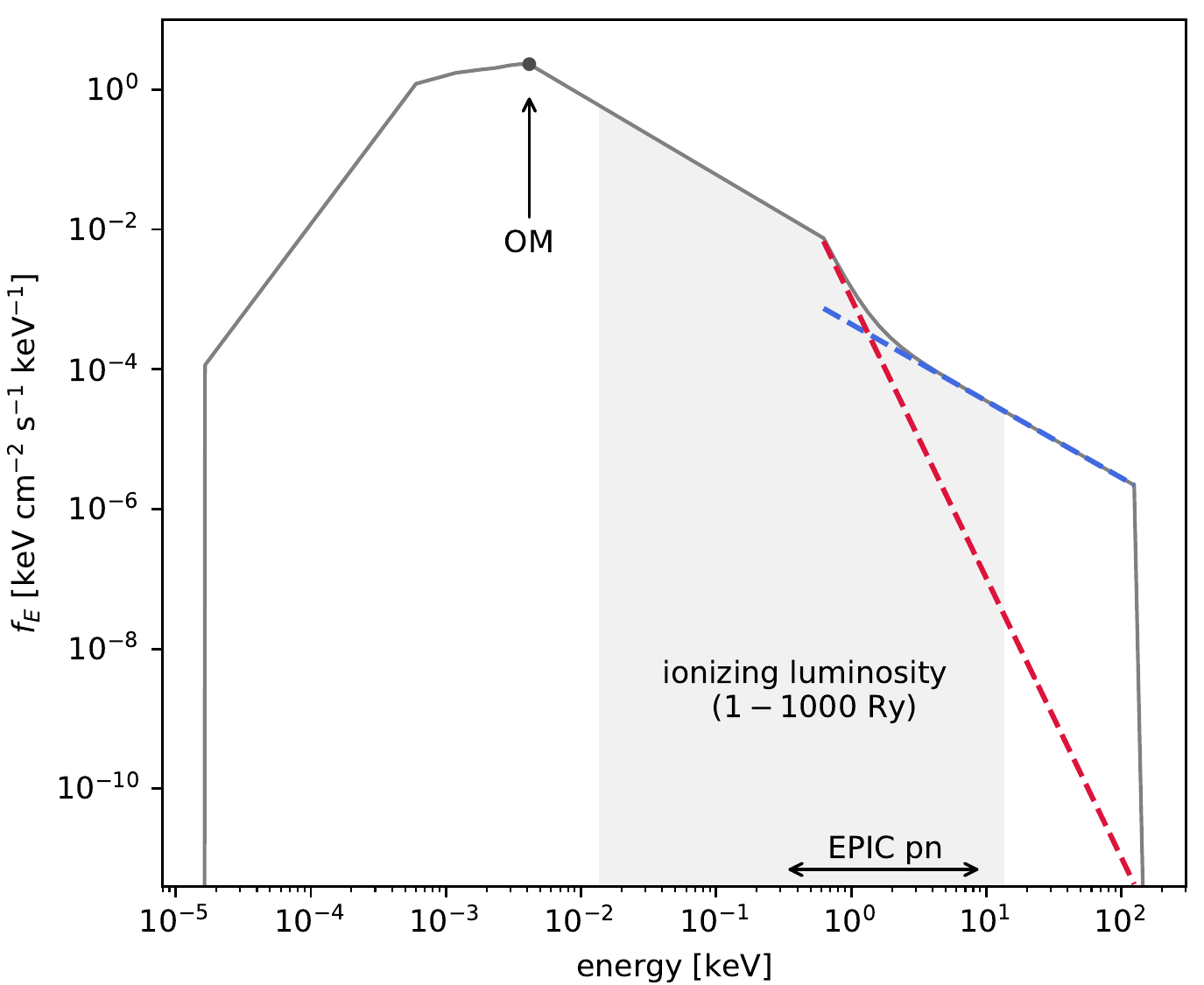}
		\caption{Average SED. The red and blue dashed lines represent power-law components whose normalisations vary simultaneously to mimic the AGN variability. The adjacent power law below $\sim 0.6~\rm keV$ pivots around the point at $ 2910$~\AA. The grey-shaded area represents the ionizing luminosity (see Section \ref{sec:concentrations} for further details).}
		\label{fig:SED}
	\end{figure}

	To model the SED of the ionizing source (see Fig. \ref{fig:SED}), a time-averaged EPIC-pn spectrum, obtained from observation 0780561701, was used, showing low variability and moderate flux \citep{Parker2017}. 
	The SED was constructed to qualitatively match the observed spectrum, while keeping the model simple to ensure that the absorber timing behaviour remains easy to isolate in the analysis. 
	
	The resulting soft X-ray part of the SED consists of a broken power law, dominant below $\sim 1~ \rm keV$, where a strong soft excess is present in the spectrum. Above the break energy at 0.6 keV, a steep decline in flux (with the photon index $\Gamma = 5.0$) occurs, while the flatter, low-energy part of this component extends down to a point obtained from the flux at $ 2910$ \AA, measured by the OM filter UVW1. This flux was corrected for extinction \citep[$E(B-V)=0.062$;][]{Schlafly2011}, using a Galactic extinction curve \citep{Cardelli1989} with $R_V=3.1$. Since the variability observed in the UV band is much smaller relative to that of the X-rays \citep{Buisson2018}, we used the time-averaged value of the UV flux. Above $\sim 1~ \rm keV$, the dominant component is a flatter ($\Gamma = 2.1$) power law, which mimics a complex reflection spectrum \citep[][]{Jiang2018}, with a cut-off at $150~\rm keV$. For the low-energy part of the SED below $ 2910$ \AA, the default AGN continuum in \textsc{cloudy} \citep{Mathews1987} was used. 
	
	During the simulations, we let the SED shape vary in time to simulate the source variability. This is, for simplicity, achieved by changing the normalisation of the flatter $\Gamma = 2.1$ power law (blue dashed line in Fig. \ref{fig:SED}) and of the steep $\Gamma = 5.0$ part of the soft power law (red dashed line). The low-energy part of this component, below the break at 0.6 keV, is fixed at the UV point at $ 2910$ \AA, around which it pivots, producing gradually decreasing variability amplitude towards lower energies. 
	
	\begin{figure*}
		\centering
		\includegraphics[width=\linewidth]{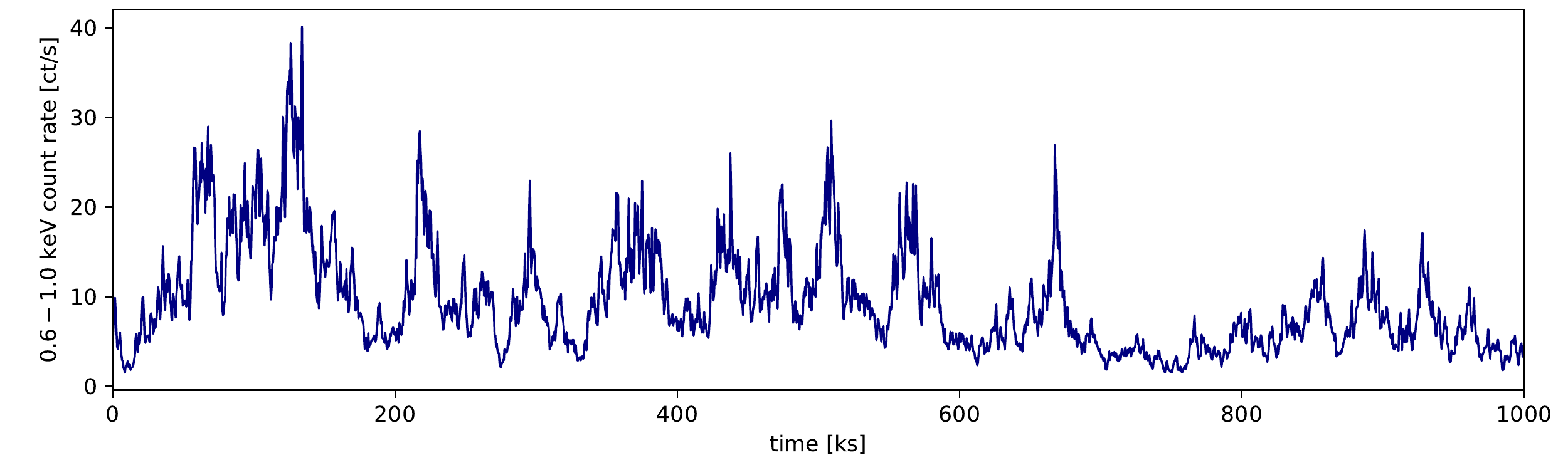}
		\caption{Simulated continuum 0.6-1.0 keV light curve, based on the X-ray variability of IRAS 13224-3809.}
		\label{fig:LC}
	\end{figure*}
	
	\subsection{Source light curve}\label{sec:srcLC}
	
	To model the interaction of the unobscured AGN radiation with the intervening outflow, we simulated a 1\,Ms light curve representative of the highly variable IRAS 13224-3809. For this purpose, we employed an algorithm presented by \citet{TimmerKoenig1995}, with which linear time series can be generated, having the desired power spectral density as provided on input.
	
	The functional form of the power spectral density used here was modelled as a broken power-law, following \citet{Summons2007}, further modified by an additional parameter $s$ which controls the sharpness of the bend. The resulting form is
	
	\begin{equation}\label{eq:bpow}
		\displaystyle P(\nu) \propto \frac{(\nu/\nu_{\rm B})^{\alpha_{\rm L}}}{[1+(\nu/\nu_{\rm B})^{(\alpha_{\rm L}-\alpha_{\rm H})s}]^{1/s}}.
	\end{equation}
	
	\noindent The break frequency $ \nu_{\rm B}$ in our model is $2\times 10^{-5}~\rm Hz$, while $\alpha_{\rm L}$, $\alpha_{\rm H}$ and $s$ are set to -1, -1.7 and 5, respectively, qualitatively matching the observed variability properties of IRAS 13224-3809. The simulated light curve, with a time step of 500 s, was then adjusted to yield a log-normal flux distribution, as observed in real data \citep[see][]{Uttley2005}. To illustrate the resulting source flux behaviour, we present the $0.6-1.0~\rm keV$ band light curve in Fig. \ref{fig:LC}. In this energy range, the SED shape is constant in time and its normalization changes according to the simulated variations.
	
	\subsection{Time-dependent ion concentrations} \label{sec:concentrations}
	
	\begin{figure}
		\centering
		\includegraphics[width=\columnwidth]{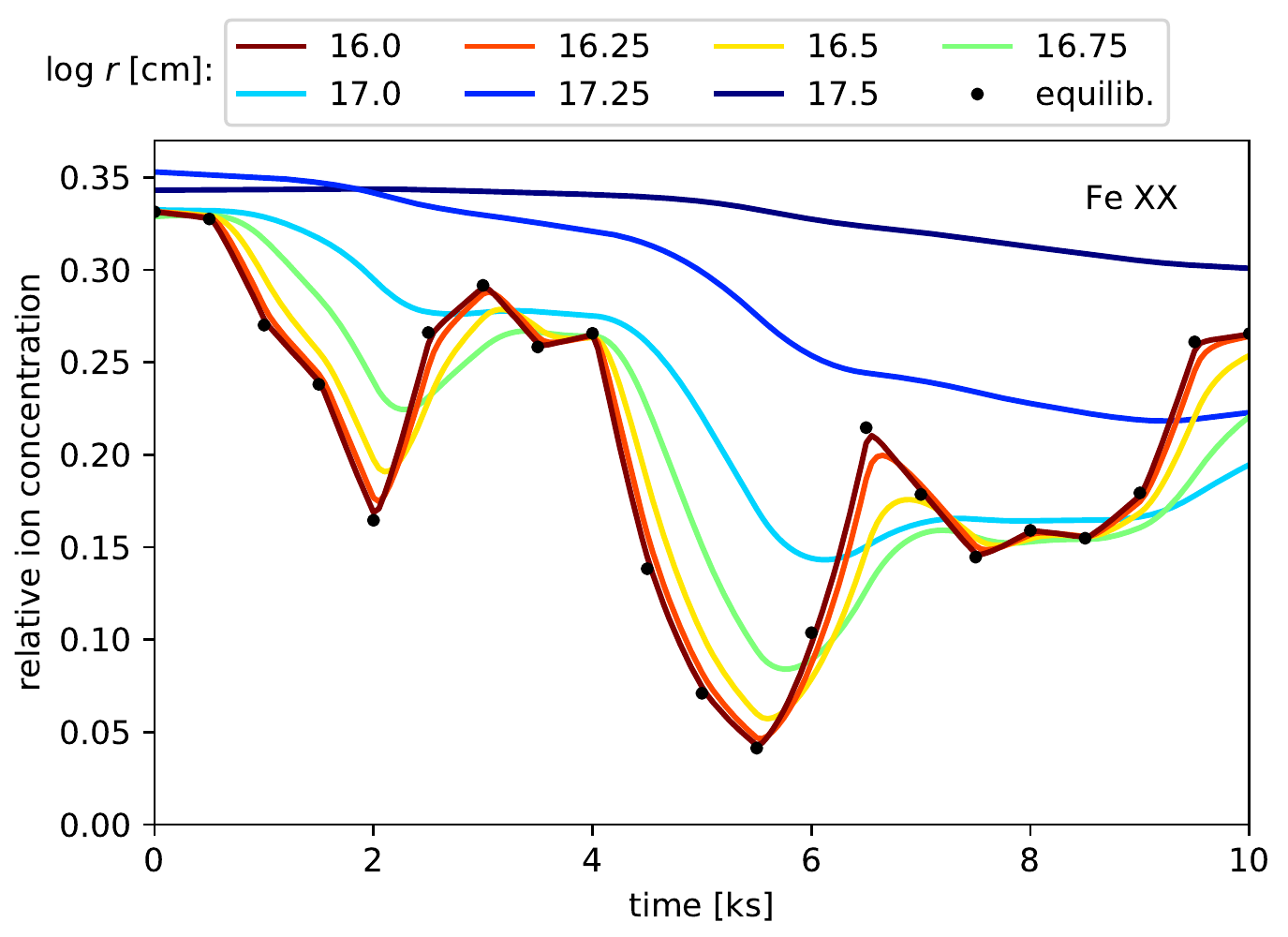}
		\caption{Time-dependent relative concentrations of \ion{Fe}{xx} simulated for several distances from the ionizing source (see legend). With increasing distance, the response becomes weaker and increasingly delayed with respect to the equilibrium concentrations instantaneous response (black dots).}
		\label{fig:timecon}
	\end{figure}
	
	Following the approach of \citet{KrolikKrisss1995}, the concentration of a certain ionization state $i$ of an element $\mathrm{X}$ as a function of time $n_{\mathrm{X}^{i}}$ can be described by a set of ionization balance equations
	
	\begin{equation}\label{eq:timecon}
		\displaystyle \frac{\dif n_{\mathrm{X}^{i}}}{\dif t} = - n_{\mathrm{e}} n_{\mathrm{X}^{i}} \alpha_{\mathrm{rec},\mathrm{X}^{i-1}} - n_{\mathrm{X}^{i}} I_{\mathrm{X}^i} + n_{\mathrm{e}} n_{\mathrm{X}^{i+1}} \alpha_{\mathrm{rec,X}^{i}} + n_{\mathrm{X}^{i-1}} I_{\mathrm{X}^{i-1}},
	\end{equation}
	
	\noindent where $n_{\mathrm{e}}$ is the electron density and $I_{\mathrm{X}^{i}}$ and $\alpha_{\mathrm{rec,X}^{i}}$ are the ionization and recombination rates, respectively, between state $i$ and $i+1$. We note that taking only photoionization and radiative recombination into account, Auger ionization, collisional ionization and three-body recombination are neglected.
	
	The equilibrium ionization and recombination rates of the outflowing gas were determined using \textsc{cloudy} \citep[version 17.01;][]{Ferland2017}, following the procedure presented by \citet[][]{Silva2016}. In the case of immediate reaction of the gas to changes of the incoming radiation, its ionization parameter scales linearly with the ionizing luminosity, given by the relation
	
	\begin{equation}\label{eq:xi}
		\xi = \frac{L_{\rm ion}}{nr^2},
	\end{equation}
	
	\noindent calculated using cgs units. The ionizing luminosity $L_{\rm ion}$ is defined in the energy range $ 1-1000~\rm Ry $, $n$ stands for the hydrogen number density and $r$ for the distance to the ionizing source. For the light curve used here, $ \log \xi $ departs from the mean value by no more than approximately $ \pm 0.44\,\rm dex$, following eq. (\ref{eq:xi}). The rates were calculated for the whole light curve with a smaller time step of $50~\rm s$, using interpolated values of $\xi$ on input, to ensure a successful integration, as detailed below. Furthermore, we assume that the denominator in eq.~(\ref{eq:xi}) remains constant over the time-scale covered in our simulations. 
	
	For the absorber, a hydrogen column density of $10^{23}~ \rm cm^{-2}$ and a mean $ \log \xi = 3.6$ were first used, similar to the outflow properties reported for IRAS~13224-3809 \citep{Jiang2018}. The metallicity was assumed to be solar with abundances from \citet{Lodders2009}. In addition to this highly ionized gas, we also probed lower ionization components: $ \log \xi = 2.6$ and $ \log \xi = 1.6$, with column densities of $2.5\times 10^{22}~ \rm cm^{-2}$ and $8\times 10^{21}~ \rm cm^{-2}$, respectively. The column densities were chosen to yield comparable total line opacity over the observable X-ray band with respect to the highly ionized case. 
	
	We solve the system of equations for a set of distances to the ionizing source, tied to the gas density by relation (\ref{eq:xi}), and assuming a constant electron density \citep[e.g.][]{Nicastro1999a, Kaastra2012, Silva2016}. The range of distances probed is $10^{15.25}-10^{19.5} \rm cm$, with a step of 0.25 dex, which, for all three average outflow ionizations, covers a broad range of gas behaviour in response to continuum variations, from immediate to no response. For the initial conditions, the concentrations were set to equilibrium values. To ensure successful integration of the stiff ordinary differential equations, we use a fourth-order Rosenbrock method with an adaptive stepper \textsc{stiff} implementation presented in \citep{Press1992}. As the original time step of the light curve (500 s) did not lead to accurate solutions for nearly-equilibrium conditions, a smaller (50 s) step was used, linearly interpolating between the points of the original light curve.
	
	An illustrative example of the evolution of the ion concentrations, namely \ion{Fe}{xx} during a small part of the light curve, is shown in Fig.~\ref{fig:timecon}. The non-equilibrium concentrations are plotted for seven distances from the ionizing source, along with equilibrium values (instantaneous response, black dots), for comparison. As the distance increases (i.e. the density decreases), the response of the gas to the continuum variations becomes delayed with respect to the equilibrium values and, concurrently, less sensitive to high-frequency variations. The amplitude of the changes decreases and eventually the ion concentration remains constant, set by the average radiation field. 
	
	\subsection{Simulated observations}
	
	Having the time-dependent ionic concentrations at our disposal, the spectra of the evolving system, when observed with \textit{Athena} X-IFU, could be simulated. In addition to the AGN model constructed in Section \ref{sec:params}, a \textsc{spex} \citep[version 3.04;][]{SPEX} component \textit{slab} \citep{Kaastra2002}, was used to represent the outflowing gas, as this model allows for the ionic column densities to be set independently. For simplicity, we assume that the gas is not affected by a systemic outflow velocity. We assumed a Gaussian velocity broadening of the absorption spectrum, of $\sigma_v = 1000~\rm km~s^{-1}$. The effect of the line broadening on the timing products is discussed in Section \ref{sec:discussion}. 
	
	\section{Results}\label{sec:timing}
	
	The analysis of the timing properties of the simulated system follows the procedure described in detail in \citet{Uttley2014}. The light curves were extracted in bins of width $\Delta E = 2.5~\rm eV$ to match the X-IFU resolution, in the energy range 0.35 -- 10 keV and were all compared to a reference, virtually unabsorbed light curve, extracted from 0.2 to 0.35 keV. To reduce the dependence of the derived timing properties on individual realisations of the underlying variability process, the 1 Ms simulation was divided into 20 segments, which were used to obtain averaged intermediate timing products \citep[for further details see Sect. 2 in][]{Uttley2014}. As a consequence of the reduced length of individual segments (50 ks), the timing products presented here probe temporal frequencies in the range $2 \times 10^{-5} - 10^{-3}~\rm Hz $. For illustration of the dependence on frequency, the results below are shown for individual Fourier frequencies resulting from the timing analysis, sampled with a constant step of $2 \times 10^{-5}~\rm Hz $.
	
	In the following, we explore how the coherence and time lag evolve as a function of energy, for three different energetics of the outflow (Section \ref{sec:concentrations}) as a function of the distance of the absorber from the source. In this example, the SED varies in a simple manner (see Section \ref{sec:params}). The effect of a more complex change of the SED along the light curve is then addressed in Section \ref{sec:varSED}.
	
	\subsection{Baseline SED}\label{sec:3xi}
	
	\subsubsection{Coherence}\label{sec:coherence}
	\begin{figure*}
		\centering
		\includegraphics[width=\linewidth]{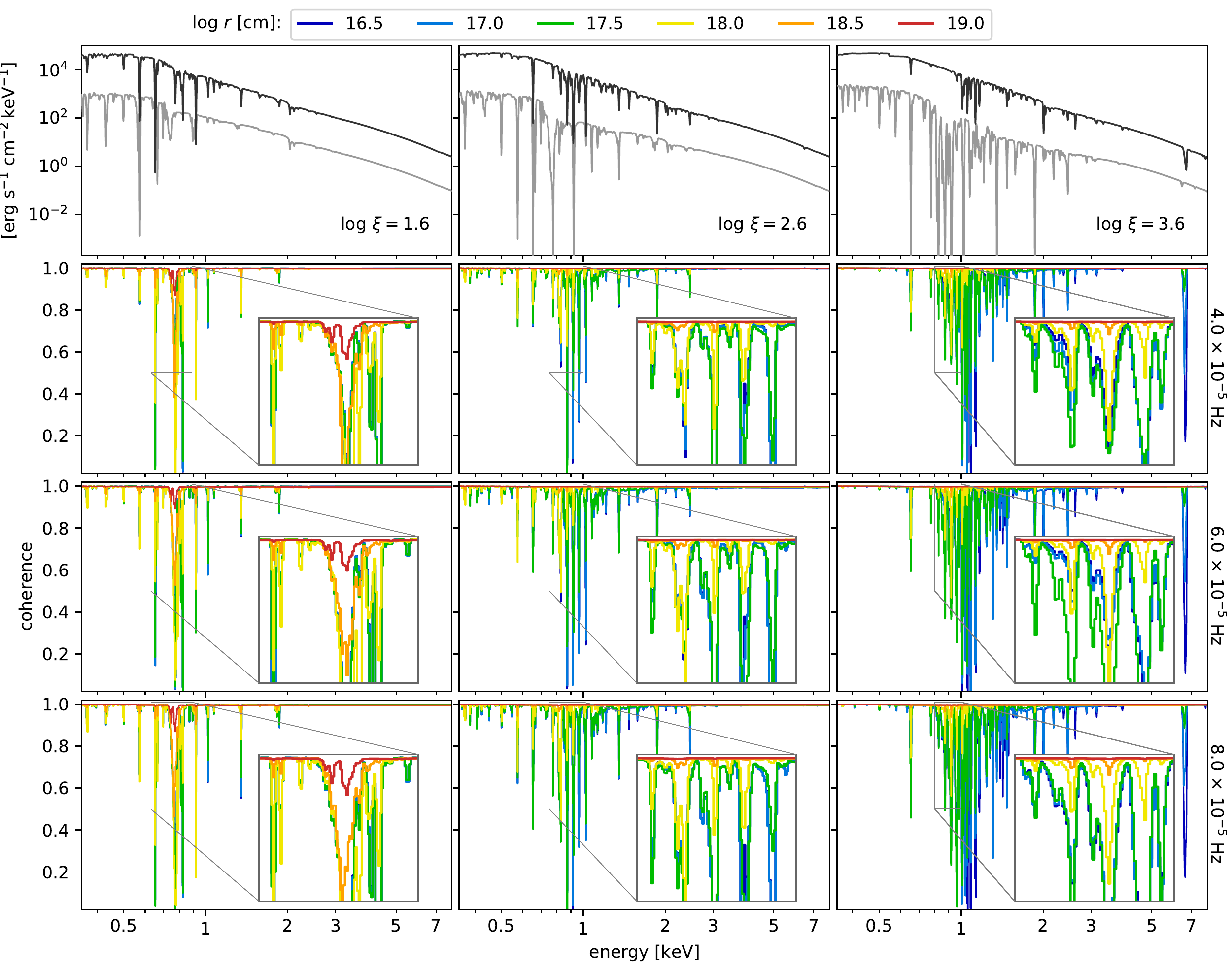}
		\caption{Top row: For every outflow, spectra of the maximum and minimum continuum flux and equilibrium ion concentrations are displayed for reference. Each simulated outflow is displayed in a different row, from low to high ionization (left to right). 2$^{\rm nd}$-- 4$^{\rm th}$ row: the coherence spectrum is displayed for six selected distances. In every row, the coherence is shown for a different frequency to illustrate the time-scale dependent behaviour in different energy bins.}
		\label{fig:coh3f3xi}
	\end{figure*}
	
	Since the response of the absorbing gas to variations of the ionizing flux is typically non-linear \citep[e.g.][]{RybickiLightman}, the light curves affected by absorption lines associated with the outflow will be generally less coherent with the unabsorbed reference light curve. In Fig. \ref{fig:coh3f3xi}, the coherence is plotted as a function of energy for all three simulated outflow energetics, displayed per column. 
	
	The top row panels contain example \textit{Athena} spectra for minimum and maximum continuum flux, as set by the light curve, and ionic concentrations of an equilibrium absorber. This is illustrative of the extent of changes in both gas ionization and continuum flux. The shape of the SED, typical of a NLS1, enhances the presence of iron ions from the M-, L- or K-shell, depending on $\xi$. The contribution of other elements potentially appearing at energies $\lesssim 3~\rm keV$, such as C, O or Ne is much smaller, as these elements are effectively depleted of their electrons by the strong radiation below $\sim 2~\rm keV$ \citep{Nicastro1999b}.
	
	In the nine panels below, the coherence spectra are displayed for 6 distances in each panel, while every row covers a different frequency. We display the frequencies (corresponding to time-scales from 25 down to 12.5 ks) in a range related to more pronounced absorption timing features. This is also in agreement with the frequencies identified in e.g. \citet[][]{Kara2016} at which lags could be influenced by absorption. The distances are selected to cover a wide range of possible gas response, from nearly-immediate (at $10^{16.5}\,\rm cm$) to only a small change in concentrations (at $10^{19.0}\,\rm cm$).
	
    For each drop in coherence, a spectral feature is visible, but not vice versa (e.g. below 0.6 keV in the right column). This happens because, for a fixed distance, the magnitude of the drop depends both on the non-linearity of the absorber response in a given energy bin and the magnitude of this response, relative to the underlying continuum component. Thus, if the line is weak or the response is close to linear, only a small drop is present. We note that the continuum variation is fully coherent, by construction of the simulation.  
	
	The magnitude of absorption-related coherence pattern does not vary dramatically as a function of frequency. The individual features, however, show moderate differences across time-scales, as can be seen by comparing the coherence spectra in different rows of Fig. \ref{fig:coh3f3xi}.
	Consequently, information from some frequencies may be more sensitive to changing distance, clearly visible in the insets of Fig. \ref{fig:coh3f3xi}.
	
	As the average ionization becomes higher, the range of distances at which the gas response is unique moves to smaller distances, as a result of longer recombination time-scale \citep{Nicastro1999a}. This effect can be seen in the insets of Fig. \ref{fig:coh3f3xi}, comparing the coherence patterns between different ionizations. While the patterns at $r = 10^{16.5}-10^{17.5}\,\rm cm$ fully overlap for the low-ionization absorber (left), the pattern at $r = 10^{17.5}\,\rm cm$ can still be distinguished in the middle panel. Finally, in the case of the highly-ionized outflow, also the pattern for $r = 10^{17.0}\,\rm cm$ shows noticeable differences when compared to that for $r = 10^{16.5}\,\rm cm$.

	\begin{figure}
		\centering
		\includegraphics[width=\linewidth]{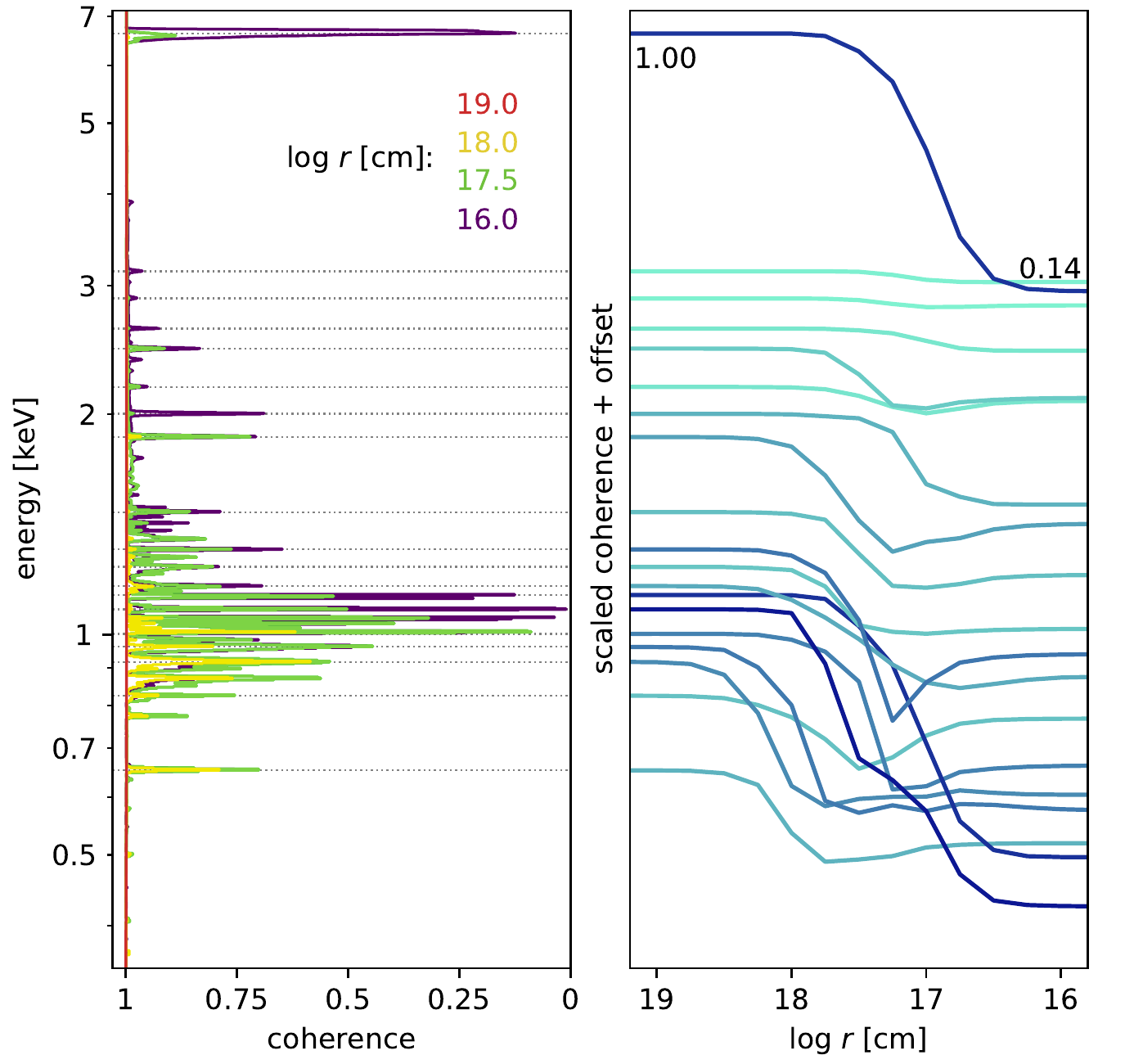}
		\caption{ The dependence of the coherence on the outflow distance, shown here for the highly ionized ($ \log \xi = 3.6 $) outflow at $4\times 10^{-5}~ \rm Hz$. Left panel: the coherence spectra (see Fig. \ref{fig:coh3f3xi}) are plotted vertically for four reference distances. The positions of some representative absorption features are marked with dotted lines, to guide the eye. Right panel: the coherence in the energy bins highlighted in the left panel is displayed this time as a function of distance, vertically offset to match the position of the corresponding dotted lines in the left panel. To enhance the readability of the plot, lines showing larger change in coherence are darker. The minimum and maximum values of the coherence for the feature at $\sim$6.5 keV are given in the plot, for reference.}
		\label{fig:cohdlines}
	\end{figure}
	
	An alternative visualization of the coherence as a function of the outflow distance is shown for the highly ionized outflow in the right panel of Fig. \ref{fig:cohdlines}. There, each line represents the coherence in one energy bin, corresponding to absorption features marked on the left, selected to illustrate the coherence behaviour. The lines are vertically offset to help identify them in the coherence-energy plot on the left, where the coherence is displayed for several distances for reference (see caption for more details). The generally decreasing coherence with decreasing distance to the ionizing source is due to increased sensitivity of this absorber to the continuum variations. This trend is halted when the equilibrium concentrations are reached in a time shorter than the continuum variability time scale, in this case at distances below $10^{16.5}~\rm cm$. Furthermore, it is clearly visible that each ion shows its own delayed response over a slightly different range of distances (or gas densities). This naturally reflects in the coherence features; the level of the average gas ionization sets not only the energies at which the drops can be observed, but also the range of distances at which the gas leaves a characteristic imprint in the coherence spectrum.

	\subsubsection{Time lags}\label{sec:timelags}
	
	\begin{figure*}
		\centering
		\includegraphics[width=\linewidth]{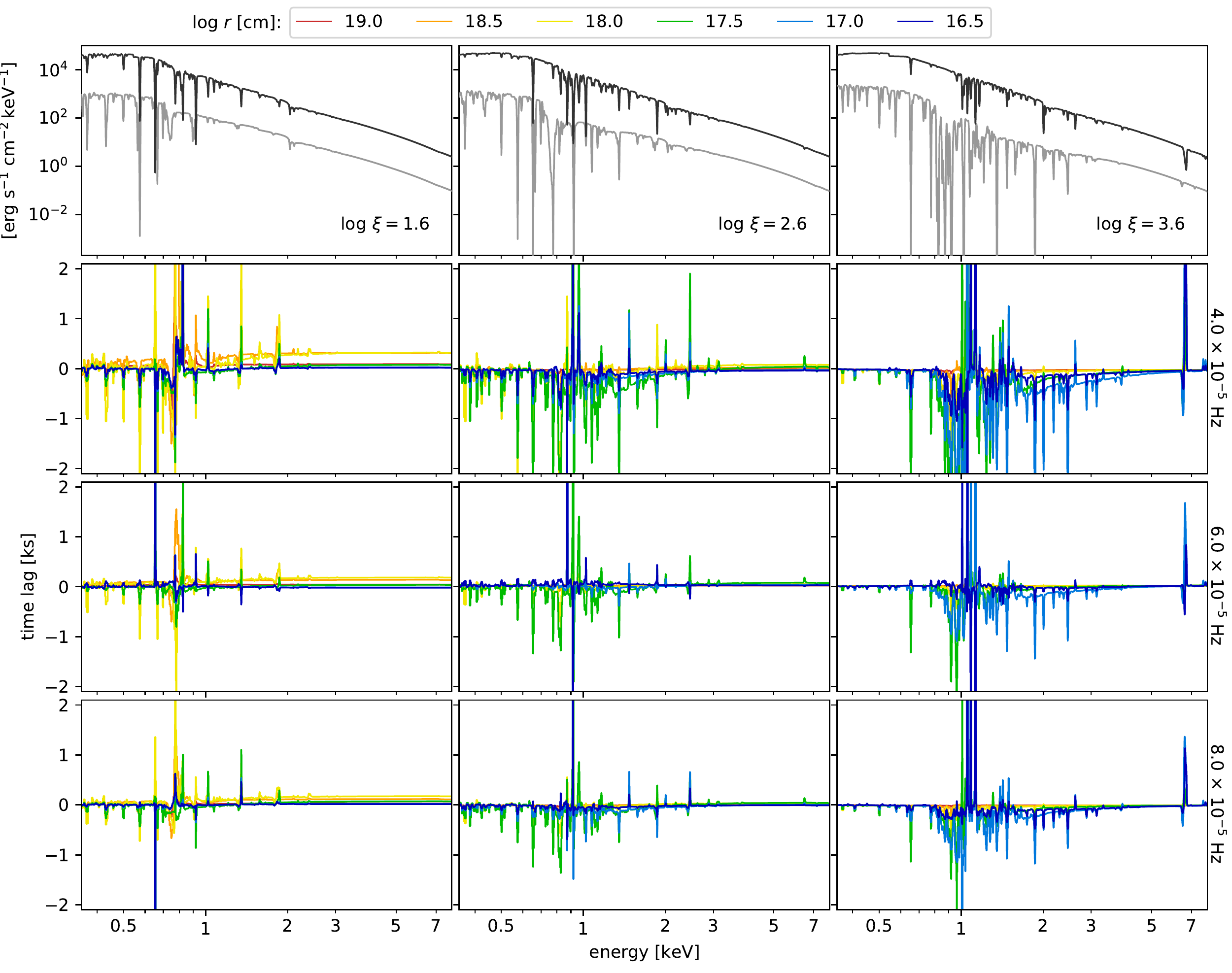}
		\caption{Top row: For every outflow, spectra of the maximum and minimum continuum flux and equilibrium ion concentrations are displayed for reference. Each simulated outflow is displayed in a different row, from low to high ionization (left to right). 2$^{\rm nd}$-- 4$^{\rm th}$ row: the lag spectrum is displayed for six selected distances. In every row, the lag is shown for a different frequency to illustrate the time-scale dependent behaviour at different energies. Where the response is non-linear, the derived time lags can reach values (exceeding the displayed range in some cases) unrelated to the physical lags (see section \ref{sec:3xi} for more details). From the definition, the negative value corresponds to the band-of-interest light curve lagging behind the reference one.}
		\label{fig:tau3f3xi}
	\end{figure*}
	
	The lag-energy spectra for the simulated outflows (Fig. \ref{fig:tau3f3xi}) are displayed for the same six distances and Fourier frequencies as in the coherence spectra in Fig. \ref{fig:coh3f3xi}, (discussed in Sect. \ref{sec:coherence}). Each column is reserved for a different simulated outflow, with increasing ionization from left to right. The time lags shown below the example spectra are displayed by row for 4, 6 and $8 \times 10^{-5}\,\rm Hz$, respectively, to illustrate the sensitivity of the absorption-related lags on increasing frequency. The lag is computed to yield negative values when the band-of-interest light curve is lagging behind the $0.2-0.35~\rm keV$ reference light curve.
	
	Unlike the coherence features, the presence of \textit{physical} time lags is limited to outflows where the gas is out of equilibrium with the ionizing radiation. In such cases, the response time of the absorbing medium translates into changes in equivalent widths (EWs) of absorption lines, which act as a delayed component in the absorbed light curve. Thus, for a fully coherent signal, time lags would only be detected in the case of a delayed response, provided sufficient absorption variability.
	
	However, due to the intrinsically incoherent signal resulting from the variable absorption, the lags produced by the outflow have a non-intuitive pattern. At larger distances, the coherence (see Fig. \ref{fig:cohdlines}) of the absorber-affected light curves is higher and the measured time lags reflect also the physical delays in the signal (see e.g. the lag features around 1 keV in the right column of Fig. \ref{fig:tau3f3xi}). Overall, the magnitude of the absorber time lags decreases with increasing temporal frequency, as can be seen comparing the lag patterns in each column.
	
	In all nine panels of Fig. \ref{fig:tau3f3xi} displaying the time lags it is noticeable that not all features show a negative lag indicative of a delayed response. In fact, the derived time lags may show an opposite sign than expected. This happens when the EW of the absorption feature is anti-correlated with the continuum flux; the absorbed light curve then seemingly leads the ionizing flux variations \citep[see also][]{Zoghbi2019}. Such behaviour is clearly visible in our simulations e.g. around 6.5 keV in the right panels of Fig. \ref{fig:tau3f3xi}. In addition, the positive lags in the unabsorbed continuum, best visible in the left panels, are unrelated to this behaviour, as the positive-sign delay is produced by lagging absorption lines in the reference band.
	
\begin{figure}
		\centering
		\includegraphics[width=\linewidth]{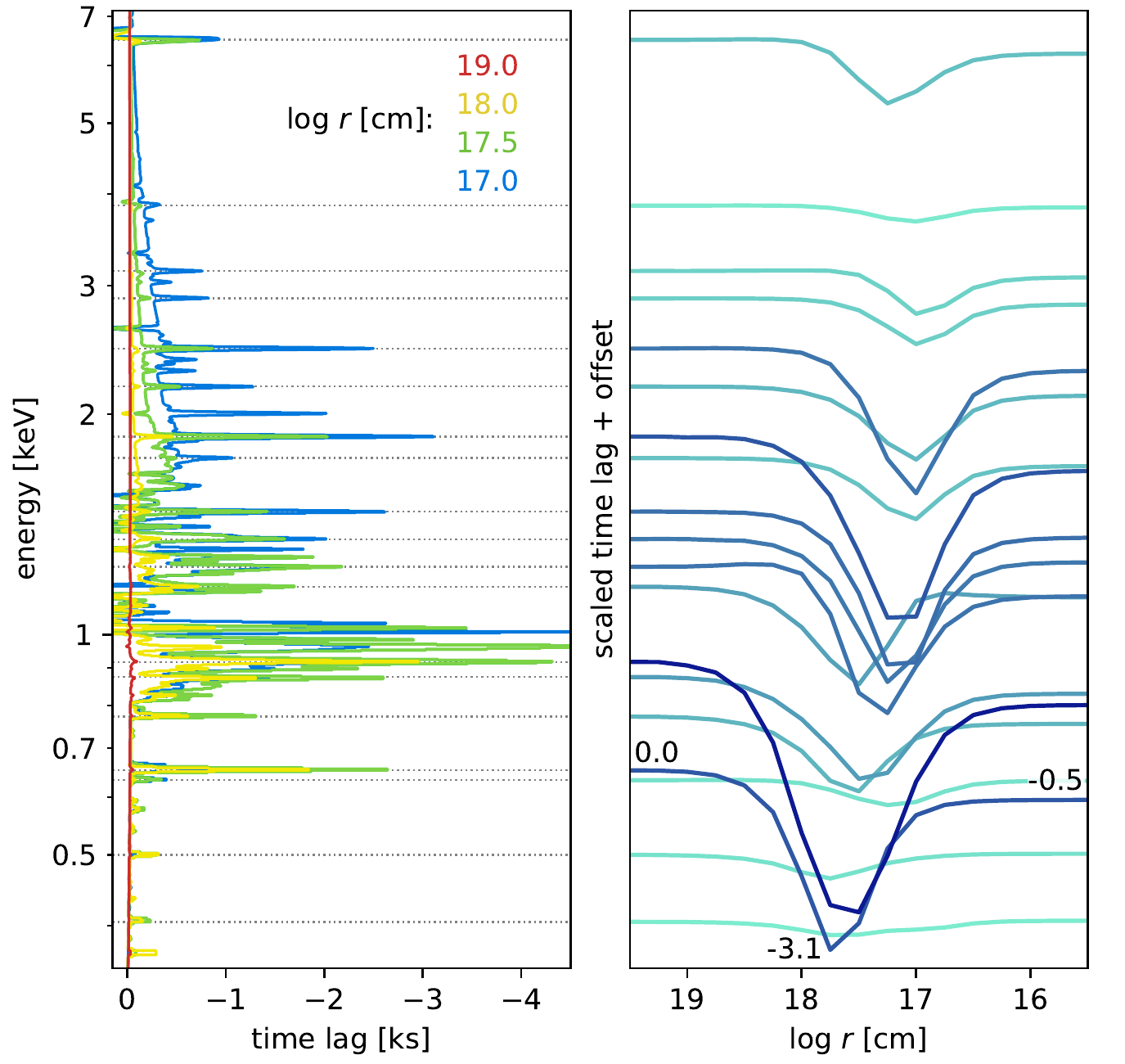}
		\caption{ The dependence of the time lag on the outflow distance, shown for the highly ionized ($ \log \xi = 3.6 $) outflow at $4\times 10^{-5}~ \rm Hz$. Left panel: the time-lag spectra (see Fig. \ref{fig:tau3f3xi}) are plotted vertically for four reference distances. The positions of some representative absorption features (with coherence greater than 0.5) are marked with dotted lines, to guide the eye. Right panel: the time lag in the energy bins highlighted in the left panel is displayed here as a function of distance, vertically offset to match the position of the corresponding dotted lines in the left panel. To enhance the readability of the plot, lines showing larger time lag are darker. For reference, the lag minimum as well as the values at $ r = 10^{15.5}$ and $10^{19.5}~\rm cm $ for the strong feature at 0.65 keV are given in the plot (in $10^3~\rm s$).}
		\label{fig:taudlines}
	\end{figure}

	A detailed illustration of the distance-dependent behaviour of the time lags is in Fig. \ref{fig:taudlines}. In the left panel, we show again, for a given frequency ($4\times 10^{-5}~ \rm Hz$) and ionization parameter ($\log \xi = 3.6$), the lag spectrum at four selected distances. In the right panel, the time lag of several representative absorption features is shown, with a vertical offset to match the position of the given band on the left. To focus on the behaviour of the physical time lags, the absorption features selected for the right panel have coherence exceeding 0.5 at all probed radii, where the contribution of artificial lags caused by incoherent signal does not dominate the radial lag profile. Consequently, the strongest coherence features (see Fig. \ref{fig:cohdlines}), in many cases showing also large positive lags at some distances, are excluded here.
	
	Following the lines in the right panel of Fig. \ref{fig:taudlines} in the direction of increasing distance (from right to left), the initial constant behaviour corresponds to an instantaneous response. Any lag visible there is caused by partial non-linearity of the gas response. After the gas ionic concentrations start departing from equilibrium values, the time lags increase in magnitude with increasing distance, since the gas can no longer follow the continuum variability immediately. Beyond the maximal detected delay, the continuum variations are too fast for the absorbing medium to respond strongly enough for the lag to remain visible, and the lag magnitude follows a decreasing trend. Finally, the gas state becomes set by the time-average radiation field and no lags are observed. 
	
	Naturally, the timing properties in the case of absorber-induced delayed response are limited to energies where the EWs of absorption lines are substantially changing. Thus, if the lines are well-separable from the continuum, the source timing behaviour can be constrained independently. 
	
	\subsection{Effects of intrinsic continuum lags}\label{sec:varSED}

\begin{figure}
		\centering
		\includegraphics[width=\linewidth]{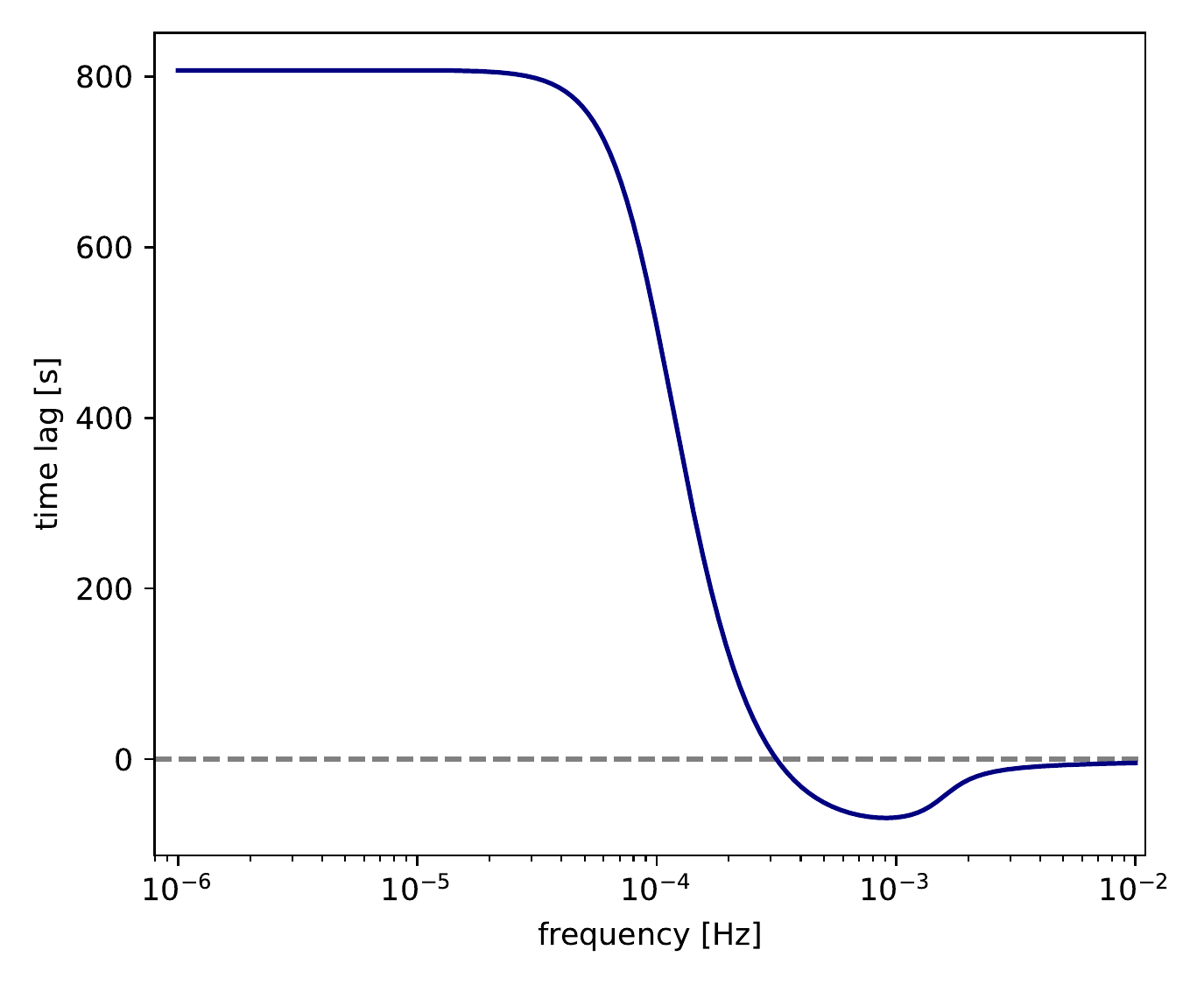}
		\caption{Time lag as a function of Fourier frequency used for modelling a time-variable shape of the ionizing continuum. The positive lag at low frequencies represents the so-called hard lag, i.e. the soft-band light curve variations precede those in the hard band. Above $\sim 3\times 10^{-4}~\rm Hz$, the value transitions to a negative (soft) lag.}
		\label{fig:cont_lags}
	\end{figure}

	To simulate the outflow timing behaviour in a more realistic system, we introduced time lags to the ionizing radiation itself. Their phenomenological prescription has the form presented in Fig.~\ref{fig:cont_lags}, mimicking the soft and hard lags observed in AGN, typically between $0.3-1.0$ and $1.0-5~\rm keV$ \citep[e.g.][]{DeMarco2013}. This function was used to relate the normalizations of the two power-law components describing a part of the SED (Fig. \ref{fig:SED}), yielding a mutually lagging continuum in the X-ray band.
	 
	At frequencies below $\sim 3\times 10^{-4}~\rm Hz$, the steep soft power-law component precedes the hard, with a maximum lag of 800~s. At higher frequencies, the sign of the time lag changes to mimic the soft lag, with a maximum delay of $\sim 70~\rm s$ at about $10^{-3}~\rm Hz$. As discussed above, the lags produced by the absorbing medium extend out to the highest frequencies, despite decreasing in magnitude significantly. At $\sim 10^{-3}~\rm Hz$, the absorber lags produced in our simulations are of about 20~s in the most delayed cases.
	
	\begin{figure}
		\centering
		\includegraphics[width=\linewidth]{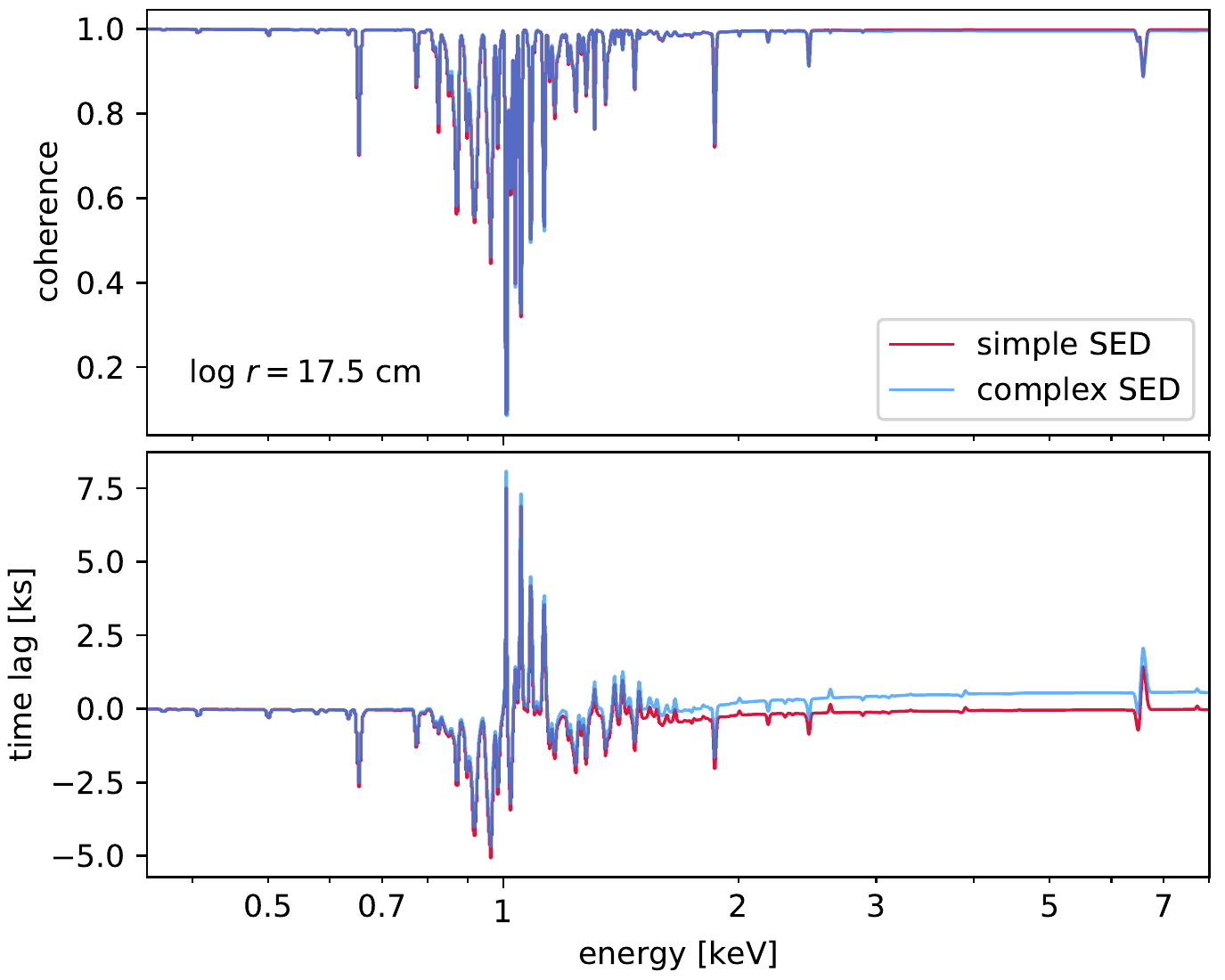}
		\caption{Comparison of simulation results for a system with the $ \log \xi = 3.6 $ outflow and $r = 10^{17.5}~ \rm cm$ and synchronously varying SED (red line) and SED composed of mutually lagging components (blue line). The coherence and time lag are shown for $4 \times 10^{-5}~ \rm Hz$ where the continuum at $E > 1~ \rm keV$ leads the soft X-ray variations.}
		\label{fig:SED_timing}
	\end{figure}
	
	Time lags in the case of absorber response are discrete features in lag spectra and if the lines are well-separable from the continuum, the source timing behaviour outside absorber-affected energies can be constrained independently. As can be seen in the bottom panel of Fig. \ref{fig:SED_timing}, the absorber lags are, however, affected by the different SED shape resulting from the lagging components. Nevertheless, the effect of continuum lags is rather small, owing to the magnitude of the lags with respect to the variability time-scale and the overall shape of the SED. Specifically, the dominant contribution of the soft ($<1~\rm keV$) part of the SED ionizing luminosity has considerably stronger effect on the gas ionic concentrations than the high-energy power law. The effect on the coherence pattern (the top panel of Fig. \ref{fig:SED_timing}) is even smaller, as the SED variability, which drives the non-linear absorption behaviour, remains close to the simultaneously variable case.

	\subsection{Observational noise}\label{sec:noise}
	The results presented thus far do not have the effects of Poisson noise taken into account. While the requirements for the signal-to-noise ratio in a given time step are lower compared to time-resolved spectroscopy, the observed count rate of the light curves still naturally limits the precision of the timing analysis results. As a consequence, suitable binning in frequency and energy might be necessary.
	
	The mean fluxes in the soft ($0.5-2$ keV) and hard ($2-10$ keV) band of the source simulated here are approximately $2\times10^{-12}~\rm erg\,s^{-1}\,cm^{-2}$ and $5\times10^{-13}~\rm erg\,s^{-1}\,cm^{-2}$, respectively, assuming the distance of IRAS~13224-3809. Consequently, the respective count rates obtained with \textit{Athena} X-IFU would be 10 and 0.3 counts per second.To include these observational limitations in our simulations, Poisson noise was applied to the simulated 500 s spectra used for light curve extraction.

\begin{figure}
		\centering
		\includegraphics[width=\linewidth]{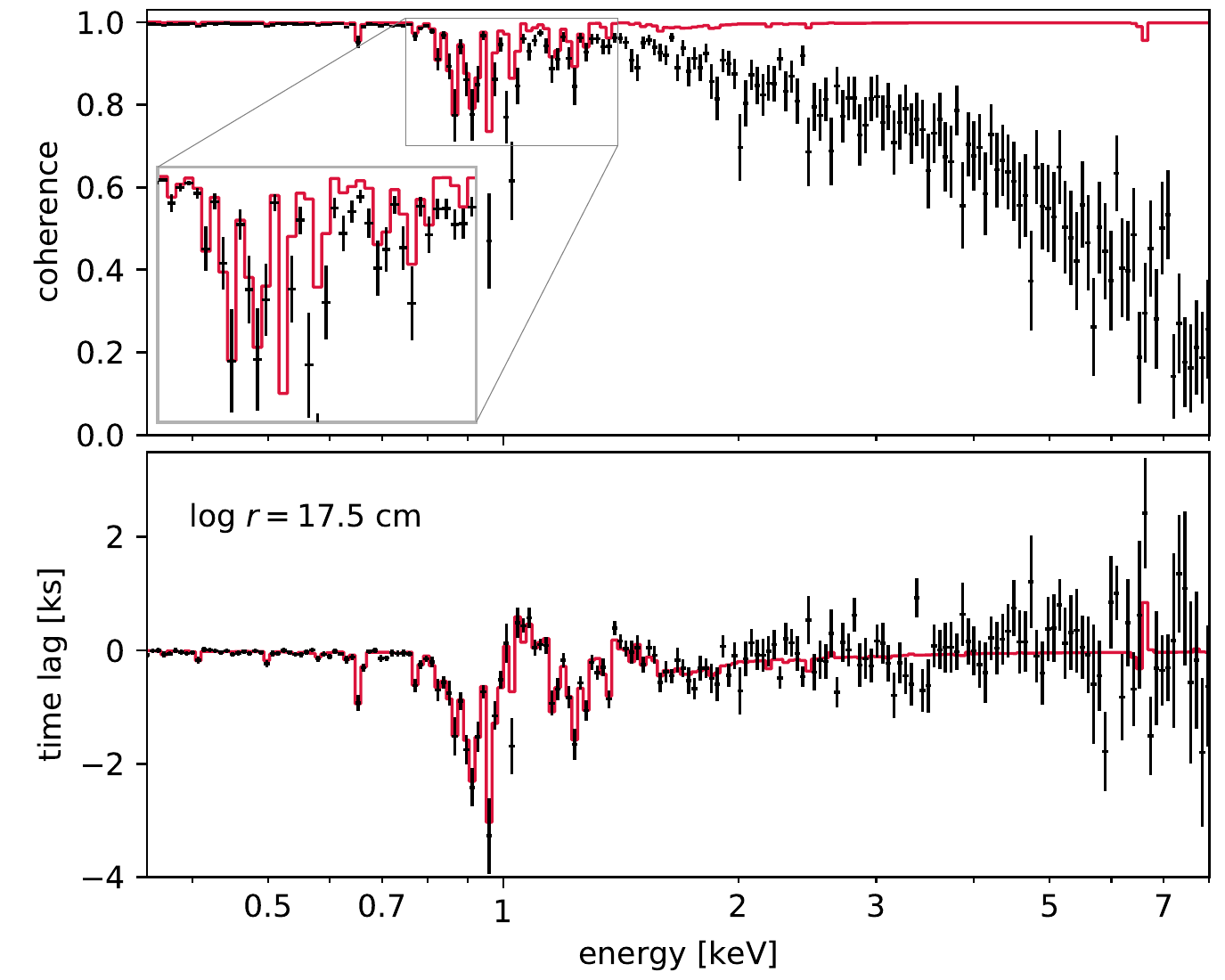}
		\caption{Coherence and time lag spectra unaffected by Poisson noise (red line) and results for simulated data  with Poisson noise applied prior the timing analysis. The results are plotted for the $ \log \xi = 3.6 $ outflow at $r = 10^{17.5}~ \rm cm$ and a frequency of $4 \times 10^{-5}~ \rm Hz$, derived from the 1~Ms simulation using the frequency width of $2 \times 10^{-5}~ \rm Hz$ for the cross-spectrum. The declining trend in coherence at high energies is caused by the lower AGN flux and consequently stronger contribution of Poisson noise to timing behaviour of these light curves.}
		\label{fig:model_noise}
	\end{figure}

	The effects of observational noise are illustrated in Fig. \ref{fig:model_noise}, where the noise-affected coherence displayed in the upper panel is the raw coherence \citep[eq. 11 in][]{Uttley2014}, computed directly from the simulated data. The results are presented for the 1~Ms simulation, averaging over 50~ks long light curve segments and using energy bins equally spaced on a logarithmic scale with $\log \Delta E \approx 0.4 $, where $E$ is in keV, to improve the signal to noise ratio.
	
	It is clearly visible that a decline in continuum flux and, consequently, an increasing influence of Poisson noise is manifested by a gradually decreasing coherence towards higher energies. The same behaviour can be seen also in strongly absorbed bins around 1 keV, where the intrinsically incoherent counting noise lowers the coherence either together with the non-linear absorber response or alone. This effect, possibly limiting the use of the outflow spectral-timing features at higher energies even with data from \textit{Athena}, can be further amplified if the absorption features are strongly blue-shifted and therefore act at energies with lower continuum flux; e.g. the feature at $\sim 6.5~\rm keV$ would shift by 1 keV for an outflow at 0.15~$c$. 

	Since the estimation of the error on the time lag is closely related to the value of the raw coherence \citep[eq. 12 in][and references therein]{Uttley2014}, the precision of the lag estimation decreases substantially both towards higher energies and in less coherent bins due to the non-linear absorber response.
	
	\section{Discussion}\label{sec:discussion} 

	\subsection{Outflow distance determination}
	
	\begin{figure*}
		\centering
		\includegraphics[width=\linewidth]{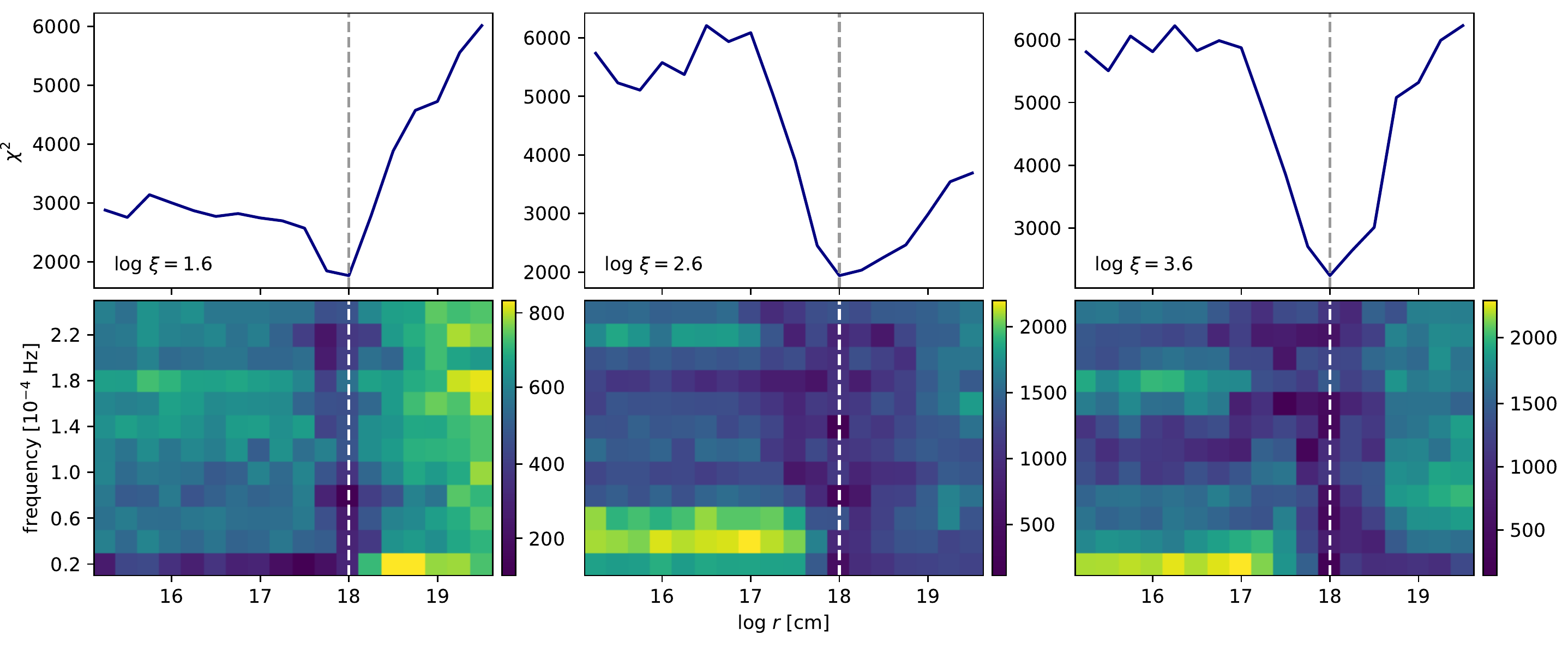}
		\caption{Example of a fit of the coherence spectrum. Top: $\chi^2$ of the coherence spectra for the models using energy in the range (0.35, 2.0) keV and first 12 frequencies, calculated with 1259 degrees of freedom. Bottom: The same but plotted for each frequency separately, with 104 degrees of freedom each. The `models' were compared to simulated data for an outflow at $r =10^{18}~\rm cm$, marked with the dashed vertical line. As can be seen in the bottom panel, while the coherence at the longest time-scales shows a strong dependence on radius, combining the information from different frequencies further improves the constraint.}
		\label{fig:fitting}
	\end{figure*}
	
	The timing properties of the photo-ionized outflows' absorption features depend on the gas density, or equivalently the distance to the ionizing source. Due to the non-linearity of the gas response to the changes in the ionizing radiation, the observed coherence of the light curves affected by variable absorption is generally lower than in unabsorbed bands, in the absence of other uncorrelated components in the signal. 
	Furthermore, while the imprint in the coherence spectra is generally dependent on the temporal frequency, all time-scales are affected, as long as the absorber ionization changes due to the source flux variations. 
	
	In order to take advantage of this behaviour to determine the outflow properties, the absorber-induced timing behaviour needs to be constrained accurately. This will be possible with the high spectral resolution of \textit{Athena} X-IFU, allowing the absorption lines and the associated timing features to be isolated from the surrounding emission spectrum. Additionally, our simulations suggest that the coherence features tend to be only marginally affected when mutually lagging continuum components are introduced, as shown in Fig. \ref{fig:SED_timing}. This is due to the low sensitivity of the SED variability itself to the delays of the individual continuum components, which, in turn, forms the absorption behaviour. As a result, absorption-induced coherence features may be easily visible also with a multi-component varying continuum.

	Considering the above-mentioned properties, the coherence spectra could be used to constrain the distance to the ionizing source. An illustrative example of the applicability of this approach is presented in Fig. \ref{fig:fitting}. There, the $\chi^2$ is shown, calculated for the raw coherence of the noise-included simulations as a function of energy, assuming that for all absorbers, the distance is $r =10^{18}~\rm cm$. In the top panels, time-scales from $50~\rm ks$ to $ \sim 4~\rm ks$ (about $ 1\,\rm d - 1\,\rm h$, covered by 12 frequencies) are used to determine the $\chi^2$. Below, the values computed separately for each frequency are colour-coded and illustrate how the sensitivity to a change in the parameter varies for different temporal frequencies. Due to the increasing influence of the noise (see Sect \ref{sec:noise}, only the energy bins below $2~\rm keV$ were used, containing most of the variable absorption.
	
	As discussed in Section \ref{sec:timing}, the average level of ionization determines the range of distances at which the gas response is delayed (and leads to a unique coherence pattern). In general, the range of distances at which the gas reacts with a delay moves closer to the ionizing source with higher average gas ionization, but ultimately the sensitivity to this parameter is set by the timing signature of the resulting absorption lines. Overall, the statistic computed for the models corresponding to all other distances differs significantly from the best-fitting one. Namely, the smallest difference, between the neighbouring parameter value ($r = 10^{17.75}~\rm cm$) in the case of the lowest-ionization outflow, is $\Delta \chi^2 > 90 $. Equivalently, the gas density would be estimated with a correspondingly high accuracy. 
	
	Knowledge of the location of the absorbing gas is crucial for determination of the mass outflow rate and ultimately also the kinetic luminosity of the outflow. Other observables, on which these quantities depend, i.e. the outflow velocity and column density, can be determined from current observations, leaving the outflow distance the remaining one to be constrained. Therefore, the constraints provided by the modelling of the spectral-timing properties of these outflows can provide valuable information in assessing their role in the AGN feedback.
	
	\subsection{Current limitations and future outlook}
	
	\begin{figure}
		\centering
		\includegraphics[width=\linewidth]{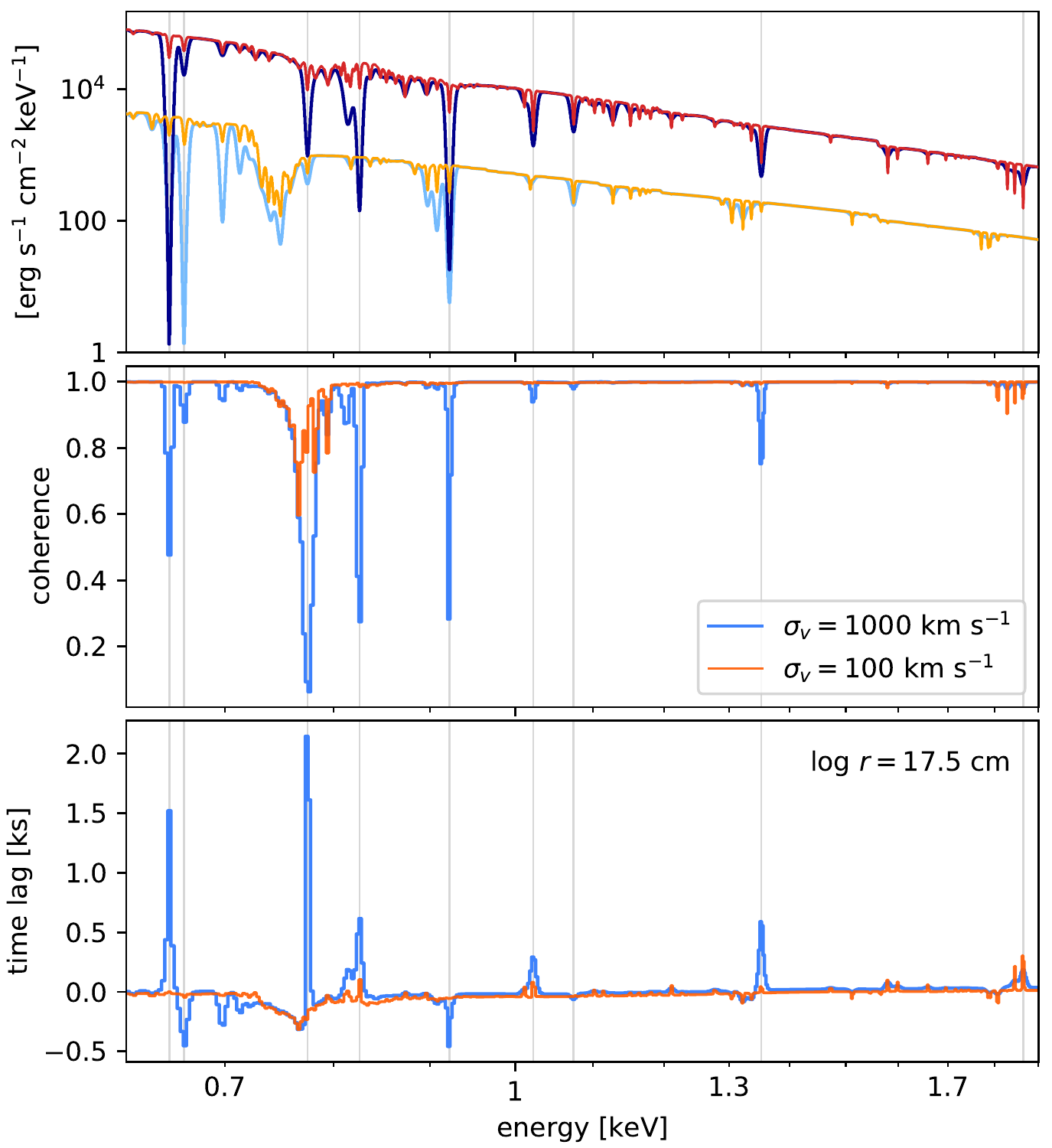}
		\caption{Top: Spectra of the outflow with the average $ \log \xi = 1.6 $, in maximum and minimum continuum flux and gas ionization, displayed for $\sigma_v = 100~\rm km~s^{-1}$ (red and yellow) and $\sigma_v = 1000~\rm km~s^{-1}$ (dark and light blue). Middle and bottom: Coherence and time lag spectra plotted for this outflow and both values of line broadening, $\sigma_v = 100~\rm km~s^{-1}$ in orange and $\sigma_v = 1000~\rm km~s^{-1}$ in blue lines. The results are plotted for a frequency of $4 \times 10^{-5}~ \rm Hz$ and $r = 10^{17.5}~ \rm cm$. Positions of the strongest lines are marked by vertical grey lines to help match the features in different panels.}
		\label{fig:sigma}
	\end{figure}
	
	The magnitude of the timing features is dependent not only on the response of the outflow spectral signatures, but also the column density and the covering factor. This has been recently shown for \textit{XMM-Newton} observations by \citet{DeMarco2020}, illustrating that the intrinsic coherence can decrease considerably, if the fraction of primary emission transmitted by the absorber in a given band is low.
	
	Furthermore, high gas column density can be associated with line saturation, which, in turn, can alter the response of the absorber-affected light curve. Fig. \ref{fig:sigma} shows an example of this effect, where the coherence and time lags are plotted for a simulation of an absorber with mean $\log \xi = 1.6 $ and turbulent line broadening of $\sigma_v = 100$ and $1000~\rm km~s^{-1}$, respectively. As can be seen in the spectra in the top panel, most of the strongest absorption features noticeable in the higher broadening scenario (blue lines) are greatly suppressed when the broadening is smaller (red and yellow), owing to a higher level of saturation. At the corresponding energies in the panels below, both the coherence dips and time lags are reduced as well. This is a consequence of a reduction of the response strength when the lines become saturated, allowing more of the coherent continuum flux to be transmitted. While the turbulent line broadening in AGN outflows usually cannot be reliably determined with the current instruments, the spectral resolution of X-IFU will allow an analysis of individual spectral lines and thus setting this parameter of the simulation accurately.
	
\begin{figure}
		\centering
		\includegraphics[width=\linewidth]{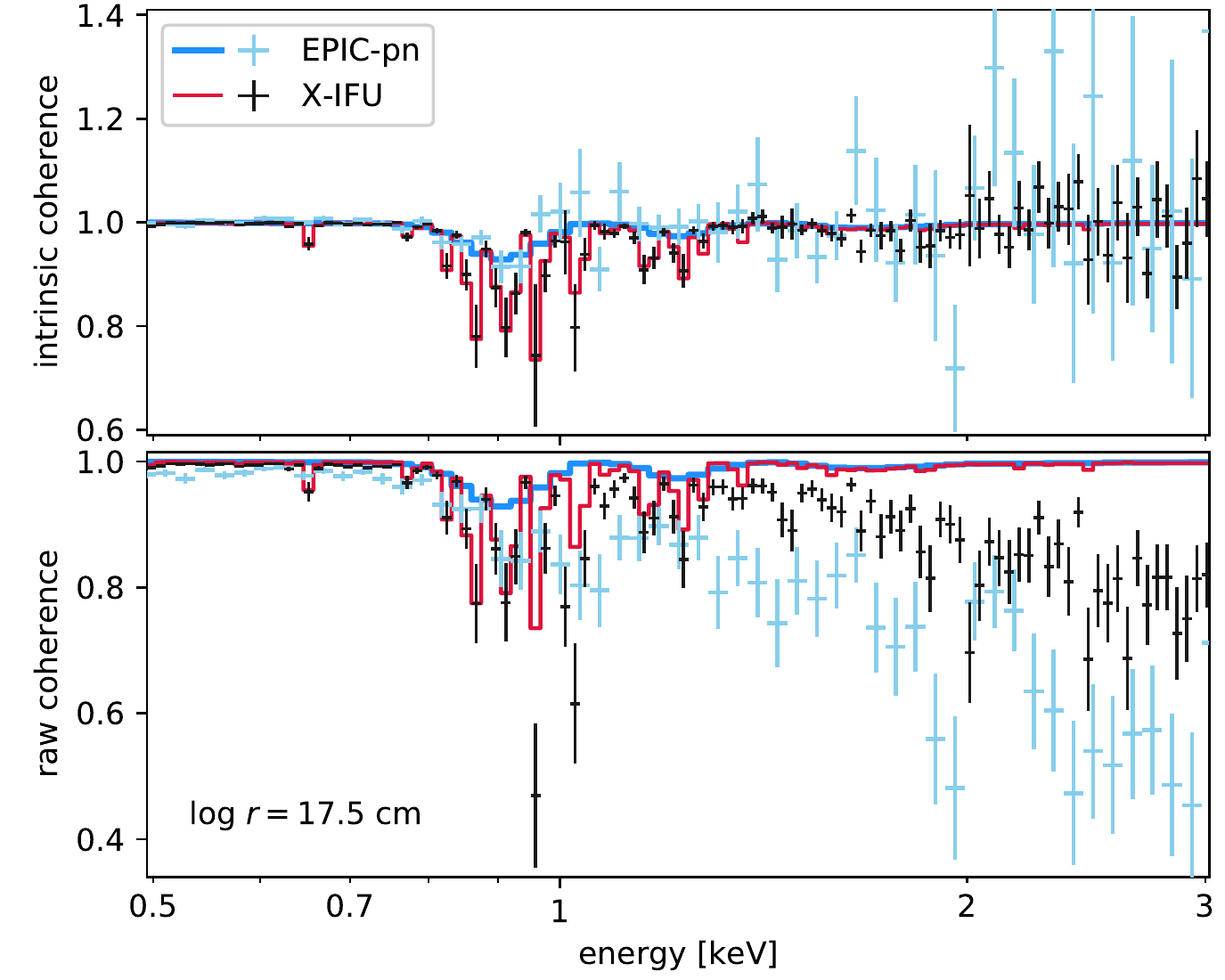}
		\caption{Intrinsic (top) and raw coherence (bottom) of the simulated source with the $ \log \xi = 3.6 $ outflow when observed with \textit{XMM-Newton} EPIC-pn and \textit{Athena} X-IFU, for comparison. The data points represent results from noise-affected light curves, whereas the solid lines show the coherence without the contribution of noise, and are thus the same in both panels.The results are derived from the 1~Ms simulation, with the frequency width of $2 \times 10^{-5}~ \rm Hz$ used for the cross-spectrum, here plotted for $r = 10^{17.5}~ \rm cm$ and $4 \times 10^{-5}~ \rm Hz$.}
		\label{fig:XMM}
	\end{figure}

	High energy resolution together with a large effective area of the instrument is needed for the outflow properties determination, as illustrated in Fig. \ref{fig:XMM}. There, the intrinsic source coherence (top panel) is estimated from the noise-included (observed) raw coherence (bottom panel) for both X-IFU and \textit{XMM-Newton} EPIC-pn detectors, for comparison. The intrinsic coherence was recovered using the approach described in eq. (8) of \citet{Vaughan1997}. As the lines illustrate, the energy resolution of EPIC-pn reduces the intrinsic coherence features, which cannot be reliably estimated from the observed data (blue points) due to the high contribution of Poisson noise, dominant already below 1 keV.
	
	Even with future instruments, the observational noise may limit the use of high-energy absorption features, such as the Fe lines in the $\sim 6.4-7~\rm keV$ range, prominent in high-ionization outflows. This is also the case for the source simulated here, resembling IRAS~13224-3809 (Sect. \ref{sec:noise}). We note, however, that this particular source is not among the brightest in the X-ray band, with an average $0.5-10$ keV flux of $1.9\times10^{-12}~\rm erg\,s^{-1}\,cm^{-2} $ \citep{Bianchi2009}. Nevertheless, depending on the source, the spectral timing analysis of future X-IFU observations may still rely on the absorption signatures present at lower energies, likely exhibiting a more complex source behaviour.
	
	The observing capabilities of upcoming X-ray missions, including \textit{Athena}, will provide new insight into the nature of the AGN emission sources in the soft X-ray band. It is likely that the observed features, connected to the absorption from AGN outflows, will be mixed with currently unresolved components, such as emission lines produced by disc reflection \citep{Barret2019}, requiring a careful approach in both data analysis and modelling. However, the rising computational power and development of simulation techniques, aiming at detailed modelling of the AGN central components geometry and dynamics and their effects on the timing properties \citep[e.g.][]{Alston2020}, will enable self-consistent modelling of the photo-ionized absorbing medium along the line of sight.
	
	The timing studies done with \textit{Athena} X-IFU observations will allow time-scales of up to about $100~\rm ks$ to be probed \citep{Barret2018}, separated by cryostat cooling periods, for longer observations. The resulting gaps, together with other instrumental effects affecting timing analysis will require a specific approach to mitigate the associated biases \citep{Huppenkothen2021}. 
	
	\section{Conclusions}\label{sec:conclusions}
	
	In this study, we simulated a set of photo-ionized AGN outflows with differing ionization, focusing on their response to the ionizing radiation of a highly variable NLS1, as they could be observed in absorption by X-IFU on board \textit{Athena}. The resulting synthetic observations were analysed employing Fourier timing techniques, concentrating in particular on the coherence and time-lag as a function of energy. 
	
	Three outflow energetics were considered, spanning from moderate to high average ionization, $\log \xi = 1.6, 2.6, 3.6$, respectively.
	
	For a range of the gas densities (or the distance to the source), the radiative processes happen on longer time-scales relative to the source variability, giving rise to a delayed and smoothed variable absorption. Due to the non-linear nature of gas response to the source radiation, the absorption features will produce local dips in coherence spectra, different depending on the frequency and the gas properties.
	
	The capability of X-IFU will allow us to exploit the coherence as a tool to determine the gas density, and equivalently its distance from the ionizing source. Our simulations show that for a typical NLS1, the location of the absorber, important for evaluating the impact of the gas on the surrounding medium, can be determined with at least the accuracy of the parameter space grid step, 0.25~dex.

    The simulations also reveal a complex behaviour of the lag spectrum. Typically, time lags can be observed even when the changes in the gas ionic concentrations are effectively instantaneous, or can have the magnitude and/or sign unrelated to the intrinsic delays in the signal. This results from the generally non-linear gas response and the combination of signal from the transmitted emission and the variable absorption in a given light curve.

    In conclusion, \textit{Athena} X-IFU, in conjunction with new analysis methods, promises an important step forward in the understanding of the physical characteristics of the AGN ionized outflows.
	
	\section*{Acknowledgements}
	AJ is partially supported by the Netherlands Organisation for Scientific Research (NWO) through the Aspasia grant connected to the Innovational Research Incentives Scheme Vidi grant 016.143.312. The Space Research Organisation of the Netherlands is financially supported by NWO. Based on observations obtained with \textit{XMM-Newton}, an European Space Agency (ESA) science mission with instruments and contributions directly funded by ESA Member States and National Aeronautics Space Administration (NASA). This research has made use of the NASA/IPAC Extragalactic Database (NED), which is funded by NASA and operated by the California Institute of Technology. We thank the anonymous referee for the prompt and constructive feedback which helped to improve the manuscript.
	
	\section*{Data availability}
	The data underlying this article will be made available in a reproduction package uploaded to Zenodo.
	
	
	
	\bibliographystyle{mnras}
	\bibliography{./ms}

\begin{thebibliography}{}
\makeatletter
\relax
\def\mn@urlcharsother{\let\do\@makeother \do\$\do\&\do\#\do\^\do\_\do\%\do\~}
\def\mn@doi{\begingroup\mn@urlcharsother \@ifnextchar [ {\mn@doi@}
  {\mn@doi@[]}}
\def\mn@doi@[#1]#2{\def\@tempa{#1}\ifx\@tempa\@empty \href
  {http://dx.doi.org/#2} {doi:#2}\else \href {http://dx.doi.org/#2} {#1}\fi
  \endgroup}
\def\mn@eprint#1#2{\mn@eprint@#1:#2::\@nil}
\def\mn@eprint@arXiv#1{\href {http://arxiv.org/abs/#1} {{\tt arXiv:#1}}}
\def\mn@eprint@dblp#1{\href {http://dblp.uni-trier.de/rec/bibtex/#1.xml}
  {dblp:#1}}
\def\mn@eprint@#1:#2:#3:#4\@nil{\def\@tempa {#1}\def\@tempb {#2}\def\@tempc
  {#3}\ifx \@tempc \@empty \let \@tempc \@tempb \let \@tempb \@tempa \fi \ifx
  \@tempb \@empty \def\@tempb {arXiv}\fi \@ifundefined
  {mn@eprint@\@tempb}{\@tempb:\@tempc}{\expandafter \expandafter \csname
  mn@eprint@\@tempb\endcsname \expandafter{\@tempc}}}

\bibitem[\protect\citeauthoryear{{Alston} et~al.,}{{Alston}
  et~al.}{2020}]{Alston2020}
{Alston} W.~N.,  et~al., 2020, \mn@doi [Nature Astronomy]
  {10.1038/s41550-019-1002-x}, \href
  {https://ui.adsabs.harvard.edu/abs/2020NatAs...4..597A} {4, 597}

\bibitem[\protect\citeauthoryear{{Barret} \& {Cappi}}{{Barret} \&
  {Cappi}}{2019}]{Barret2019}
{Barret} D.,  {Cappi} M.,  2019, \mn@doi [\aap] {10.1051/0004-6361/201935817},
  \href {https://ui.adsabs.harvard.edu/abs/2019A&A...628A...5B} {628, A5}

\bibitem[\protect\citeauthoryear{{Barret} et~al.,}{{Barret}
  et~al.}{2018}]{Barret2018}
{Barret} D.,  et~al., 2018, in {den Herder} J.-W.~A.,  {Nikzad} S.,
  {Nakazawa} K.,  eds,  Society of Photo-Optical Instrumentation Engineers
  (SPIE) Conference Series Vol. 10699, Space Telescopes and Instrumentation
  2018: Ultraviolet to Gamma Ray. p. 106991G (\mn@eprint {arXiv} {1807.06092}),
  \mn@doi{10.1117/12.2312409}

\bibitem[\protect\citeauthoryear{{Bianchi}, {Guainazzi}, {Matt}, {Fonseca
  Bonilla}  \& {Ponti}}{{Bianchi} et~al.}{2009}]{Bianchi2009}
{Bianchi} S.,  {Guainazzi} M.,  {Matt} G.,  {Fonseca Bonilla} N.,   {Ponti} G.,
   2009, \mn@doi [\aap] {10.1051/0004-6361:200810620}, \href
  {https://ui.adsabs.harvard.edu/abs/2009A&A...495..421B} {495, 421}

\bibitem[\protect\citeauthoryear{{Blustin}, {Page}, {Fuerst},
  {Branduardi-Raymont}  \& {Ashton}}{{Blustin} et~al.}{2005}]{Blustin2005}
{Blustin} A.~J.,  {Page} M.~J.,  {Fuerst} S.~V.,  {Branduardi-Raymont} G.,
  {Ashton} C.~E.,  2005, \mn@doi [\aap] {10.1051/0004-6361:20041775}, \href
  {https://ui.adsabs.harvard.edu/abs/2005A&A...431..111B} {431, 111}

\bibitem[\protect\citeauthoryear{{Boller}, {Brandt}  \& {Fink}}{{Boller}
  et~al.}{1996}]{Boller1996}
{Boller} T.,  {Brandt} W.~N.,   {Fink} H.,  1996, \aap, \href
  {https://ui.adsabs.harvard.edu/abs/1996A&A...305...53B} {305, 53}

\bibitem[\protect\citeauthoryear{{Buisson} et~al.,}{{Buisson}
  et~al.}{2018}]{Buisson2018}
{Buisson} D.~J.~K.,  et~al., 2018, \mn@doi [\mnras] {10.1093/mnras/sty008},
  \href {https://ui.adsabs.harvard.edu/abs/2018MNRAS.475.2306B} {475, 2306}

\bibitem[\protect\citeauthoryear{{Cappi}, {Tombesi}  \& {Giustini}}{{Cappi}
  et~al.}{2013}]{Cappi2013}
{Cappi} M.,  {Tombesi} F.,   {Giustini} M.,  2013, \memsai, \href
  {https://ui.adsabs.harvard.edu/abs/2013MmSAI..84..691C} {84, 691}

\bibitem[\protect\citeauthoryear{{Cardelli}, {Clayton}  \& {Mathis}}{{Cardelli}
  et~al.}{1989}]{Cardelli1989}
{Cardelli} J.~A.,  {Clayton} G.~C.,   {Mathis} J.~S.,  1989, \mn@doi [\apj]
  {10.1086/167900}, \href
  {https://ui.adsabs.harvard.edu/abs/1989ApJ...345..245C} {345, 245}

\bibitem[\protect\citeauthoryear{{Chartas} \& {Canas}}{{Chartas} \&
  {Canas}}{2018}]{Chartas2018}
{Chartas} G.,  {Canas} M.~H.,  2018, \mn@doi [\apj] {10.3847/1538-4357/aae438},
  \href {https://ui.adsabs.harvard.edu/abs/2018ApJ...867..103C} {867, 103}

\bibitem[\protect\citeauthoryear{{De Marco}, {Ponti}, {Cappi}, {Dadina},
  {Uttley}, {Cackett}, {Fabian}  \& {Miniutti}}{{De Marco}
  et~al.}{2013}]{DeMarco2013}
{De Marco} B.,  {Ponti} G.,  {Cappi} M.,  {Dadina} M.,  {Uttley} P.,  {Cackett}
  E.~M.,  {Fabian} A.~C.,   {Miniutti} G.,  2013, \mn@doi [\mnras]
  {10.1093/mnras/stt339}, \href
  {https://ui.adsabs.harvard.edu/abs/2013MNRAS.431.2441D} {431, 2441}

\bibitem[\protect\citeauthoryear{{De Marco} et~al.,}{{De Marco}
  et~al.}{2020}]{DeMarco2020}
{De Marco} B.,  et~al., 2020, \mn@doi [\aap] {10.1051/0004-6361/201936470},
  \href {https://ui.adsabs.harvard.edu/abs/2020A&A...634A..65D} {634, A65}

\bibitem[\protect\citeauthoryear{{Fabian} et~al.,}{{Fabian}
  et~al.}{2009}]{Fabian2009}
{Fabian} A.~C.,  et~al., 2009, \mn@doi [\nat] {10.1038/nature08007}, \href
  {https://ui.adsabs.harvard.edu/abs/2009Natur.459..540F} {459, 540}

\bibitem[\protect\citeauthoryear{{Ferland} et~al.,}{{Ferland}
  et~al.}{2017}]{Ferland2017}
{Ferland} G.~J.,  et~al., 2017, \rmxaa, \href
  {https://ui.adsabs.harvard.edu/abs/2017RMxAA..53..385F} {53, 385}

\bibitem[\protect\citeauthoryear{{Galeev}, {Rosner}  \& {Vaiana}}{{Galeev}
  et~al.}{1979}]{Galeev1979}
{Galeev} A.~A.,  {Rosner} R.,   {Vaiana} G.~S.,  1979, \mn@doi [\apj]
  {10.1086/156957}, \href
  {https://ui.adsabs.harvard.edu/abs/1979ApJ...229..318G} {229, 318}

\bibitem[\protect\citeauthoryear{{Gallo}, {Boller}, {Tanaka}, {Fabian},
  {Brandt}, {Welsh}, {Anabuki}  \& {Haba}}{{Gallo} et~al.}{2004}]{Gallo2004}
{Gallo} L.~C.,  {Boller} T.,  {Tanaka} Y.,  {Fabian} A.~C.,  {Brandt} W.~N.,
  {Welsh} W.~F.,  {Anabuki} N.,   {Haba} Y.,  2004, \mn@doi [\mnras]
  {10.1111/j.1365-2966.2004.07196.x}, \href
  {https://ui.adsabs.harvard.edu/abs/2004MNRAS.347..269G} {347, 269}

\bibitem[\protect\citeauthoryear{{George}, {Turner}, {Netzer}, {Nandra},
  {Mushotzky}  \& {Yaqoob}}{{George} et~al.}{1998}]{George1998}
{George} I.~M.,  {Turner} T.~J.,  {Netzer} H.,  {Nandra} K.,  {Mushotzky}
  R.~F.,   {Yaqoob} T.,  1998, \mn@doi [\apjs] {10.1086/313067}, \href
  {https://ui.adsabs.harvard.edu/abs/1998ApJS..114...73G} {114, 73}

\bibitem[\protect\citeauthoryear{{Haardt} \& {Maraschi}}{{Haardt} \&
  {Maraschi}}{1993}]{Haardt1993}
{Haardt} F.,  {Maraschi} L.,  1993, \mn@doi [\apj] {10.1086/173020}, \href
  {https://ui.adsabs.harvard.edu/abs/1993ApJ...413..507H} {413, 507}

\bibitem[\protect\citeauthoryear{{Huppenkothen} \& {Bachetti}}{{Huppenkothen}
  \& {Bachetti}}{2021}]{Huppenkothen2021}
{Huppenkothen} D.,  {Bachetti} M.,  2021, \mn@doi [\mnras]
  {10.1093/mnras/stab3437}, \href
  {https://ui.adsabs.harvard.edu/abs/2021MNRAS.tmp.3294H} {}

\bibitem[\protect\citeauthoryear{{Jiang} et~al.,}{{Jiang}
  et~al.}{2018}]{Jiang2018}
{Jiang} J.,  et~al., 2018, \mn@doi [\mnras] {10.1093/mnras/sty836}, \href
  {https://ui.adsabs.harvard.edu/abs/2018MNRAS.477.3711J} {477, 3711}

\bibitem[\protect\citeauthoryear{{Kaastra}, {Mewe}  \&
  {Nieuwenhuijzen}}{{Kaastra} et~al.}{1996}]{SPEX}
{Kaastra} J.~S.,  {Mewe} R.,   {Nieuwenhuijzen} H.,  1996, in {Yamashita} K.,
  {Watanabe} T.,  eds, UV and X-ray Spectroscopy of Astrophysical and
  Laboratory Plasmas. Universal Academy Press, Tokyo, pp 411--414

\bibitem[\protect\citeauthoryear{{Kaastra}, {Steenbrugge}, {Raassen}, {van der
  Meer}, {Brinkman}, {Liedahl}, {Behar}  \& {de Rosa}}{{Kaastra}
  et~al.}{2002}]{Kaastra2002}
{Kaastra} J.~S.,  {Steenbrugge} K.~C.,  {Raassen} A.~J.~J.,  {van der Meer}
  R.~L.~J.,  {Brinkman} A.~C.,  {Liedahl} D.~A.,  {Behar} E.,   {de Rosa} A.,
  2002, \mn@doi [\aap] {10.1051/0004-6361:20020235}, \href
  {https://ui.adsabs.harvard.edu/abs/2002A&A...386..427K} {386, 427}

\bibitem[\protect\citeauthoryear{{Kaastra} et~al.,}{{Kaastra}
  et~al.}{2012}]{Kaastra2012}
{Kaastra} J.~S.,  et~al., 2012, \mn@doi [\aap] {10.1051/0004-6361/201118161},
  \href {https://ui.adsabs.harvard.edu/abs/2012A&A...539A.117K} {539, A117}

\bibitem[\protect\citeauthoryear{{Kara}, {Alston}  \& {Fabian}}{{Kara}
  et~al.}{2016}]{Kara2016}
{Kara} E.,  {Alston} W.,   {Fabian} A.,  2016, \mn@doi [Astronomische
  Nachrichten] {10.1002/asna.201612332}, \href
  {https://ui.adsabs.harvard.edu/abs/2016AN....337..473K} {337, 473}

\bibitem[\protect\citeauthoryear{{Kara} et~al.,}{{Kara}
  et~al.}{2019}]{Kara2019}
{Kara} E.,  et~al., 2019, \mn@doi [\nat] {10.1038/s41586-018-0803-x}, \href
  {https://ui.adsabs.harvard.edu/abs/2019Natur.565..198K} {565, 198}

\bibitem[\protect\citeauthoryear{{Komossa} \& {Meerschweinchen}}{{Komossa} \&
  {Meerschweinchen}}{2000}]{Komossa2000}
{Komossa} S.,  {Meerschweinchen} J.,  2000, \aap, \href
  {https://ui.adsabs.harvard.edu/abs/2000A&A...354..411K} {354, 411}

\bibitem[\protect\citeauthoryear{{Krolik} \& {Kriss}}{{Krolik} \&
  {Kriss}}{1995}]{KrolikKrisss1995}
{Krolik} J.~H.,  {Kriss} G.~A.,  1995, \mn@doi [\apj] {10.1086/175896}, \href
  {https://ui.adsabs.harvard.edu/abs/1995ApJ...447..512K} {447, 512}

\bibitem[\protect\citeauthoryear{{Laha}, {Reynolds}, {Reeves}, {Kriss},
  {Guainazzi}, {Smith}, {Veilleux}  \& {Proga}}{{Laha} et~al.}{2021}]{Laha2021}
{Laha} S.,  {Reynolds} C.~S.,  {Reeves} J.,  {Kriss} G.,  {Guainazzi} M.,
  {Smith} R.,  {Veilleux} S.,   {Proga} D.,  2021, \mn@doi [Nature Astronomy]
  {10.1038/s41550-020-01255-2}, \href
  {https://ui.adsabs.harvard.edu/abs/2021NatAs...5...13L} {5, 13}

\bibitem[\protect\citeauthoryear{{Leighly}}{{Leighly}}{1999}]{Leighly1999}
{Leighly} K.~M.,  1999, \mn@doi [\apjs] {10.1086/313277}, \href
  {https://ui.adsabs.harvard.edu/abs/1999ApJS..125..297L} {125, 297}

\bibitem[\protect\citeauthoryear{{Leighly}, {Mushotzky}, {Nandra}  \&
  {Forster}}{{Leighly} et~al.}{1997}]{Leighly1997}
{Leighly} K.~M.,  {Mushotzky} R.~F.,  {Nandra} K.,   {Forster} K.,  1997,
  \mn@doi [\apjl] {10.1086/310950}, \href
  {https://ui.adsabs.harvard.edu/abs/1997ApJ...489L..25L} {489, L25}

\bibitem[\protect\citeauthoryear{{Lodders}, {Palme}  \& {Gail}}{{Lodders}
  et~al.}{2009}]{Lodders2009}
{Lodders} K.,  {Palme} H.,   {Gail} H.-P.,  2009, Landolt B{\"o}rnstein

\bibitem[\protect\citeauthoryear{{Mangham}, {Knigge}, {Matthews}, {Long}, {Sim}
   \& {Higginbottom}}{{Mangham} et~al.}{2017}]{Mangham2017}
{Mangham} S.~W.,  {Knigge} C.,  {Matthews} J.~H.,  {Long} K.~S.,  {Sim} S.~A.,
   {Higginbottom} N.,  2017, \mn@doi [\mnras] {10.1093/mnras/stx1863}, \href
  {https://ui.adsabs.harvard.edu/abs/2017MNRAS.471.4788M} {471, 4788}

\bibitem[\protect\citeauthoryear{{Mathews} \& {Ferland}}{{Mathews} \&
  {Ferland}}{1987}]{Mathews1987}
{Mathews} W.~G.,  {Ferland} G.~J.,  1987, \mn@doi [\apj] {10.1086/165843},
  \href {https://ui.adsabs.harvard.edu/abs/1987ApJ...323..456M} {323, 456}

\bibitem[\protect\citeauthoryear{{Miller}, {Turner}, {Reeves}  \&
  {Braito}}{{Miller} et~al.}{2010}]{Miller2010}
{Miller} L.,  {Turner} T.~J.,  {Reeves} J.~N.,   {Braito} V.,  2010, \mn@doi
  [\mnras] {10.1111/j.1365-2966.2010.17261.x}, \href
  {https://ui.adsabs.harvard.edu/abs/2010MNRAS.408.1928M} {408, 1928}

\bibitem[\protect\citeauthoryear{{Mizumoto}, {Ebisawa}, {Tsujimoto}, {Done},
  {Hagino}  \& {Odaka}}{{Mizumoto} et~al.}{2019}]{Mizumoto2019}
{Mizumoto} M.,  {Ebisawa} K.,  {Tsujimoto} M.,  {Done} C.,  {Hagino} K.,
  {Odaka} H.,  2019, \mn@doi [\mnras] {10.1093/mnras/sty3056}, \href
  {https://ui.adsabs.harvard.edu/abs/2019MNRAS.482.5316M} {482, 5316}

\bibitem[\protect\citeauthoryear{{Nandra} et~al.,}{{Nandra}
  et~al.}{2013}]{Nandra2013}
{Nandra} K.,  et~al., 2013, arXiv e-prints, \href
  {https://ui.adsabs.harvard.edu/abs/2013arXiv1306.2307N} {p. arXiv:1306.2307}

\bibitem[\protect\citeauthoryear{{Nicastro}, {Fiore}, {Perola}  \&
  {Elvis}}{{Nicastro} et~al.}{1999a}]{Nicastro1999a}
{Nicastro} F.,  {Fiore} F.,  {Perola} G.~C.,   {Elvis} M.,  1999a, \mn@doi
  [\apj] {10.1086/306736}, \href
  {https://ui.adsabs.harvard.edu/abs/1999ApJ...512..184N} {512, 184}

\bibitem[\protect\citeauthoryear{{Nicastro}, {Fiore}  \& {Matt}}{{Nicastro}
  et~al.}{1999b}]{Nicastro1999b}
{Nicastro} F.,  {Fiore} F.,   {Matt} G.,  1999b, \mn@doi [\apj]
  {10.1086/307187}, \href
  {https://ui.adsabs.harvard.edu/abs/1999ApJ...517..108N} {517, 108}

\bibitem[\protect\citeauthoryear{{Parker} et~al.,}{{Parker}
  et~al.}{2017}]{Parker2017}
{Parker} M.~L.,  et~al., 2017, \mn@doi [\nat] {10.1038/nature21385}, \href
  {https://ui.adsabs.harvard.edu/abs/2017Natur.543...83P} {543, 83}

\bibitem[\protect\citeauthoryear{{Ponti}, {Papadakis}, {Bianchi}, {Guainazzi},
  {Matt}, {Uttley}  \& {Bonilla}}{{Ponti} et~al.}{2012}]{Ponti2012}
{Ponti} G.,  {Papadakis} I.,  {Bianchi} S.,  {Guainazzi} M.,  {Matt} G.,
  {Uttley} P.,   {Bonilla} N.~F.,  2012, \mn@doi [\aap]
  {10.1051/0004-6361/201118326}, \href
  {https://ui.adsabs.harvard.edu/abs/2012A&A...542A..83P} {542, A83}

\bibitem[\protect\citeauthoryear{{Press}, {Teukolsky}, {Vetterling}  \&
  {Flannery}}{{Press} et~al.}{1992}]{Press1992}
{Press} W.~H.,  {Teukolsky} S.~A.,  {Vetterling} W.~T.,   {Flannery} B.~P.,
  1992, {Numerical recipes in FORTRAN. The art of scientific computing}.
Cambridge University Press

\bibitem[\protect\citeauthoryear{{Reynolds}}{{Reynolds}}{1997}]{Reynolds1997}
{Reynolds} C.~S.,  1997, \mn@doi [\mnras] {10.1093/mnras/286.3.513}, \href
  {https://ui.adsabs.harvard.edu/abs/1997MNRAS.286..513R} {286, 513}

\bibitem[\protect\citeauthoryear{{Reynolds}, {Young}, {Begelman}  \&
  {Fabian}}{{Reynolds} et~al.}{1999}]{Reynolds1999}
{Reynolds} C.~S.,  {Young} A.~J.,  {Begelman} M.~C.,   {Fabian} A.~C.,  1999,
  \mn@doi [\apj] {10.1086/306913}, \href
  {https://ui.adsabs.harvard.edu/abs/1999ApJ...514..164R} {514, 164}

\bibitem[\protect\citeauthoryear{{Ross} \& {Fabian}}{{Ross} \&
  {Fabian}}{1993}]{Ross1993}
{Ross} R.~R.,  {Fabian} A.~C.,  1993, \mn@doi [\mnras]
  {10.1093/mnras/261.1.74}, \href
  {https://ui.adsabs.harvard.edu/abs/1993MNRAS.261...74R} {261, 74}

\bibitem[\protect\citeauthoryear{Rybicki \& Lightman}{Rybicki \&
  Lightman}{1991}]{RybickiLightman}
Rybicki G.,  Lightman A.,  1991, Radiative Processes in Astrophysics.
A Wiley-Interscience publication, Wiley, \url
  {https://books.google.nl/books?id=LtdEjNABMlsC}

\bibitem[\protect\citeauthoryear{{Schlafly} \& {Finkbeiner}}{{Schlafly} \&
  {Finkbeiner}}{2011}]{Schlafly2011}
{Schlafly} E.~F.,  {Finkbeiner} D.~P.,  2011, \mn@doi [\apj]
  {10.1088/0004-637X/737/2/103}, \href
  {https://ui.adsabs.harvard.edu/abs/2011ApJ...737..103S} {737, 103}

\bibitem[\protect\citeauthoryear{{Silva}, {Uttley}  \& {Costantini}}{{Silva}
  et~al.}{2016}]{Silva2016}
{Silva} C.~V.,  {Uttley} P.,   {Costantini} E.,  2016, \mn@doi [\aap]
  {10.1051/0004-6361/201628555}, \href
  {https://ui.adsabs.harvard.edu/abs/2016A&A...596A..79S} {596, A79}

\bibitem[\protect\citeauthoryear{{Stella}}{{Stella}}{1990}]{Stella1990}
{Stella} L.,  1990, \mn@doi [\nat] {10.1038/344747a0}, \href
  {https://ui.adsabs.harvard.edu/abs/1990Natur.344..747S} {344, 747}

\bibitem[\protect\citeauthoryear{{Summons}, {Ar{\'e}valo}, {McHardy}, {Uttley}
  \& {Bhaskar}}{{Summons} et~al.}{2007}]{Summons2007}
{Summons} D.~P.,  {Ar{\'e}valo} P.,  {McHardy} I.~M.,  {Uttley} P.,   {Bhaskar}
  A.,  2007, \mn@doi [\mnras] {10.1111/j.1365-2966.2006.11797.x}, \href
  {https://ui.adsabs.harvard.edu/abs/2007MNRAS.378..649S} {378, 649}

\bibitem[\protect\citeauthoryear{{Timmer} \& {Koenig}}{{Timmer} \&
  {Koenig}}{1995}]{TimmerKoenig1995}
{Timmer} J.,  {Koenig} M.,  1995, \aap, \href
  {https://ui.adsabs.harvard.edu/abs/1995A&A...300..707T} {300, 707}

\bibitem[\protect\citeauthoryear{{Tombesi}, {Cappi}, {Reeves}, {Palumbo},
  {Yaqoob}, {Braito}  \& {Dadina}}{{Tombesi} et~al.}{2010}]{Tombesi2010}
{Tombesi} F.,  {Cappi} M.,  {Reeves} J.~N.,  {Palumbo} G.~G.~C.,  {Yaqoob} T.,
  {Braito} V.,   {Dadina} M.,  2010, \mn@doi [\aap]
  {10.1051/0004-6361/200913440}, \href
  {https://ui.adsabs.harvard.edu/abs/2010A&A...521A..57T} {521, A57}

\bibitem[\protect\citeauthoryear{{Uttley}, {McHardy}  \& {Vaughan}}{{Uttley}
  et~al.}{2005}]{Uttley2005}
{Uttley} P.,  {McHardy} I.~M.,   {Vaughan} S.,  2005, \mn@doi [\mnras]
  {10.1111/j.1365-2966.2005.08886.x}, \href
  {https://ui.adsabs.harvard.edu/abs/2005MNRAS.359..345U} {359, 345}

\bibitem[\protect\citeauthoryear{{Uttley}, {Cackett}, {Fabian}, {Kara}  \&
  {Wilkins}}{{Uttley} et~al.}{2014}]{Uttley2014}
{Uttley} P.,  {Cackett} E.~M.,  {Fabian} A.~C.,  {Kara} E.,   {Wilkins} D.~R.,
  2014, \mn@doi [\aapr] {10.1007/s00159-014-0072-0}, \href
  {https://ui.adsabs.harvard.edu/abs/2014A&ARv..22...72U} {22, 72}

\bibitem[\protect\citeauthoryear{{Vaughan} \& {Nowak}}{{Vaughan} \&
  {Nowak}}{1997}]{Vaughan1997}
{Vaughan} B.~A.,  {Nowak} M.~A.,  1997, \mn@doi [\apjl] {10.1086/310430}, \href
  {https://ui.adsabs.harvard.edu/abs/1997ApJ...474L..43V} {474, L43}

\bibitem[\protect\citeauthoryear{{Zoghbi}, {Fabian}, {Reynolds}  \&
  {Cackett}}{{Zoghbi} et~al.}{2012}]{Zoghbi2012}
{Zoghbi} A.,  {Fabian} A.~C.,  {Reynolds} C.~S.,   {Cackett} E.~M.,  2012,
  \mn@doi [\mnras] {10.1111/j.1365-2966.2012.20587.x}, \href
  {https://ui.adsabs.harvard.edu/abs/2012MNRAS.422..129Z} {422, 129}

\bibitem[\protect\citeauthoryear{{Zoghbi}, {Miller}  \& {Cackett}}{{Zoghbi}
  et~al.}{2019}]{Zoghbi2019}
{Zoghbi} A.,  {Miller} J.~M.,   {Cackett} E.,  2019, \mn@doi [\apj]
  {10.3847/1538-4357/ab3e31}, \href
  {https://ui.adsabs.harvard.edu/abs/2019ApJ...884...26Z} {884, 26}

\makeatother
\end{thebibliography}
	
	
	
	
	
	
	\bsp	
	\label{lastpage}
\end{document}